\documentclass[twocolumn,showpacs,preprintnumbers,amsmath,amssymb,prd,letterpaper,floatfix,nofootinbib,superscriptaddress]{revtex4-1}
\usepackage{graphics}
\usepackage{graphicx}
\usepackage{epsfig}
\usepackage[usenames]{color}
\usepackage{longtable}
\PassOptionsToPackage{hyphens}{url}\usepackage{hyperref}
\usepackage{breakurl}
\usepackage{latexsym}
\usepackage{amsmath}
\usepackage{amsthm}
\usepackage{amsfonts}
\usepackage{amssymb}
\usepackage{dsfont}
\usepackage{bm}

\newcommand\T{\rule{0pt}{2.6ex}}
\newcommand\B{\rule[-1.2ex]{0pt}{0pt}}
\newcommand{\FC}{\;,}
\newcommand{\FD}{\;.}
\newcommand{\I}{\mathrm{i}}  
\newcommand{\E}{\mathrm{e}}  
\newcommand{\be}{\begin{equation}}
\newcommand{\ee}{\end{equation}}
\newcommand{\bea}{\begin{eqnarray}}
\newcommand{\eea}{\end{eqnarray}}
\newcommand{\ubar}{\overline{u}}

\newcommand{\sbar}{\overline{s}}

\newcommand{\qbar}{\overline{q}}
\newcommand{\eq}[1]{(\ref{#1})}
\newcommand{\trace}{\mathop{\mathrm{tr}}\nolimits}
\newcommand{\unitmatrix}{\ensuremath{\mathbf{1}}}

\begin{document}

\title{$D_{s}$ mesons with $DK$ and $D^{*}K$ scattering near threshold}

\author{C.~B.~Lang}
\email{christian.lang@uni-graz.at}
\affiliation{Institute of Physics,  University of Graz, A--8010 Graz, Austria}

\author{Luka Leskovec}
\email{luka.leskovec@ijs.si}
\affiliation{Jozef Stefan Institute, 1000 Ljubljana, Slovenia}

\author{Daniel Mohler}
\email{dmohler@fnal.gov}
\affiliation{Fermi National Accelerator Laboratory, Batavia, Illinois 60510-5011, USA}

\author{Sasa Prelovsek}
\email{sasa.prelovsek@ijs.si}
\affiliation{Department of Physics, University of Ljubljana, 1000 Ljubljana, Slovenia}
\affiliation{Jozef Stefan Institute, 1000 Ljubljana, Slovenia}

\author{R.~M.~Woloshyn}
\email{rwww@triumf.ca}
\affiliation{TRIUMF, 4004 Wesbrook Mall Vancouver, BC V6T 2A3, Canada}

\date{\today}

\begin{abstract}
$D_s$ mesons are studied in three quantum channels ($J^P=0^+$, $1^+$ and $2^+$), where experiments have identified the very narrow $D_{s0}^*(2317)$, $D_{s1}(2460)$ and narrow $D_{s1}(2536)$, $D_{s2}^*(2573)$. We explore the effect of nearby $DK$ and  $D^*K$ thresholds on the subthreshold states using lattice QCD. Our simulation is done on two very different ensembles of gauge configurations (2 or 2+1 dynamical quarks, Pion mass of 266 or 156 MeV, lattice size $16^3\times 32$ or $32^3\times 64$). In addition to $\qbar q$ operators we also include meson-meson interpolators in the correlation functions. This clarifies the identification of the states above and below the scattering thresholds. The ensemble with $m_\pi \simeq 156~$MeV renders the $D_{s1}(2460)$ as a strong interaction bound state $44(10)~$MeV below $D^*K$ threshold, which is in agreement with the experiment.  The $D_{s0}^*(2317)$ is found $37(17)~$MeV below $DK$ threshold, close to experiment value of $45~$MeV. The narrow resonances $D_{s1}(2536)$ and $D_{s2}^*(2573)$ are also found close to the experimental masses. 
\end{abstract}


\maketitle

\section{Introduction}

During the past decade there have been significant advances in our knowledge of charmed meson spectroscopy. In the charm-strange meson sector, states consistent with the expected positive parity $D_s$ meson ground states have been observed \cite{Beringer:1900zz}. In the heavy charm quark limit \cite{Isgur:1991wq} these states can be grouped into two multiplets characterized by $j=1/2$ and $j=3/2$ where $j$ is the sum of the strange quark spin and the ($p$-wave) orbital angular momentum. Prior to their discovery the $D_{s0}^*$ and $D_{s1}$ states associated with the $j=1/2$ multiplet were expected to lie above the $DK$ and $D^*K$ thresholds respectively and to be very broad since they could then break apart into $s$-wave meson pairs.\footnote{This is what is observed in the $D$ meson sector \cite{Beringer:1900zz}. Two very broad states $0^+$ and $1^+$ decaying into $s$-wave meson pairs and two higher-lying states $1^+$ and $2^+$ which are more narrow, presumably decaying into $d$-wave pairs.} Instead, experiments found $D_{s0}^*$ and $D_{s1}$ mesons which lie close to, but below, the $DK$ and $D^*K$ thresholds, respectively. These results, combined with the surprisingly similar mass of $D_{s0}^{*}(2317)$ to its non-strange partner $D_{0}(2400)$ led to many ideas such as tetraquarks (see for example  \cite{Dmitrasinovic:2005gc}), molecular states, etc.. Particularly relevant was the suggestion \cite{vanBeveren:2003kd} that the coupling of $\bar{c}s$ to the $DK$ threshold plays an important dynamical role in lowering the mass of the physical state.

Lattice QCD provides a nonperturbative framework to calculate hadron properties and it has been applied extensively to $D_{s}$ spectroscopy \cite{Boyle:1997aq,Boyle:1997rk,Hein:2000qu,Lewis:2000sv,Bali:2003jv,diPierro:2003iw,Dougall:2003hv,Namekawa:2011wt,Mohler:2011ke,Bali:2012ua,Bali:2011dc,Moir:2013ub,Kalinowski:2013wsa,Wagner:2013laa,Mohler:2013rwa}. Early quenched lattice QCD calculations \cite{Boyle:1997aq,Boyle:1997rk,Hein:2000qu,Lewis:2000sv,Bali:2003jv,diPierro:2003iw,Dougall:2003hv} which considered $D_{s0}^*$ found energy levels in line with quark model expectations \cite{Godfrey:1985xj}, that is, substantially above the physical $DK$ threshold. Later dynamical lattice QCD simulations \cite{Namekawa:2011wt,Mohler:2011ke,Bali:2012ua,Bali:2011dc,Moir:2013ub,Kalinowski:2013wsa,Wagner:2013laa} obtained somewhat smaller $D_{s0}^*$ masses but still generally larger than the experimental value. In these simulations the $D_{s0}^*$ and $D_{s1}$ mesons were described using only quark-antiquark interpolating operators. In a recent lattice study of $D_{s0}^*(2317)$ where operators for $DK$ scattering states were included in the operator basis good agreement with the experimental value of the mass was found \cite{Mohler:2013rwa}. In Ref. \cite{Mohler:2013rwa} the mass was no longer obtained directly from the two-point meson correlation function but was inferred from the scattering phase shift using L\"uscher's finite volume method \cite{Luscher:1985dn,Luscher:1986pf,Luscher:1990ux,Luscher:1991cf}.

Separate from the comparison of the calculated mass to the experimental value is the question of the mass relative to the two meson scattering threshold. If the $j=1/2$ mesons $D_{s0}^*$ and $D_{s1}$ had masses above the $DK$ and $D^*K$ thresholds they would likely have large widths analogous to what is found for the  $j=1/2$ mesons in the D-meson sector. Physically $D_{s0}^*$ and $D_{s1}$ have masses below the physical thresholds. However, the outcome from lattice simulations seems to depend somewhat delicately on up and down quark (equivalently, Pion) mass and choice of operators. The quark mass dependence is delicate because the D-meson and Kaon have valence up and down quark content and would naturally be expected to be more sensitive to Pion mass than the $D_s$ interpolated as a $\bar{c}s$ state. Present simulations \cite{Moir:2013ub,Mohler:2013rwa} indicate that for Pion masses substantially larger than physical and using only a $\bar{c}s$ basis the $D_{s0}^*$ will appear below the $DK$ threshold. However, in a near to physical light quark mass simulation \cite{Mohler:2013rwa} the $D_{s0}^*$ was above threshold using only a $\bar{c}s$ basis.

When scattering states are included in the simulation one can, in addition to extracting masses, also calculate scattering lengths. Although these are not amenable to experimental determination for heavy-light mesons comparisons of results between different calculational approaches can be made. For the $DK$ system lattice QCD results for scattering lengths have been already presented in \cite{Liu:2012zya} and \cite{Mohler:2013rwa}. They have also been calculated in effective field theories, for example, in \cite{Geng:2010vw,Liu:2011mi,Liu:2012zya,Wu:2012ef}. The dependence on $m_\pi$ of the mass differences between the scalar and pseudoscalar heavy-light mesons  was investigated in \cite{Becirevic:2004uv}.

The phenomenological approach starts from experimental evidence and models the scattering amplitudes by various methods. Unitarized chiral expansions have been widely used. This allows then to vary the parameters and trace the reaction of bound states and/or resonances. Replacing the continuum  space integrals by discrete sums leads to discrete energy levels which then can be compared with the results of the (ab initio) lattice calculations. A chiral unitary coupled channel study claims that $D_{s0}^*$ develops dynamically from $DK$ and $D_s\eta$ \cite{MartinezTorres:2011pr}. For a dynamical coupled-channel approach for meson-meson in $s$-wave see also \cite{Doring:2011ip}.

In this paper we present simulation results for the complete set ($J^P=0^+$, $1^+$ and $2^+$) of low-lying positive parity  $D_{s}$ mesons. For the $J^P=0^+$ and $1^+$ channels the scattering method including two-meson operators in the interpolating operator basis is used. While the results for $J^P=0^+$ were presented previously \cite{Mohler:2013rwa}, here all the details of the calculation are discussed. For the $D_{s2}^*$ only quark-antiquark operators were used as it is a narrow resonance in experiment and is expected to be described well as a $\bar{c}s$ state; the same approach was used in a previous study of D mesons \cite{Mohler:2012na}.

When two-meson scattering operators are included the lattice simulation becomes quite challenging due to the presence of three and four point correlation functions. These contain Wick contractions with what we term backtracking loops (see, for example, Fig.~\ref{fig:triangle_D_s_DK}). The calculation of these terms requires quark propagators which connect different spatial points on the same lattice time slice. Since the correlation functions are needed for all lattice time distances a method that can calculate quark propagators between any pair of lattice sites is required. For this the distillation technique \cite{Peardon:2009gh} is used.
The essential idea is that quark fields are smeared with a function that can be expressed in terms of the eigenvectors of some convenient smearing operator (the 3D lattice Laplacian is used here). The eigenvectors can be contracted with the quark propagators and these so-called perambulators can be constructed and used as the basic building blocks of correlation functions with any Wick contraction.

For our small lattices we use full distillation \cite{Peardon:2009gh}. As the lattice volume is increased the number of required eigenvectors to keep the source profile roughly the same physical size becomes eventually prohibitively large. To remedy this issue for the ensemble of larger lattices used in this work the stochastic distillation variant \cite{Morningstar:2011ka} is employed. Stochastic distillation has been used previously in Refs. \cite{Morningstar:2013bda,Mohler:2013rwa,Jost:2013uba}. See also Ref. \cite{Foley:2005ac} for a more general discussion of stochastic methods.

Section \ref{sec:tools} contains a discussion of the how the calculations were carried out. Details of the gauge configurations, the distillation methods, extraction of phase shifts and so on are presented. The results for {$D_{s0}^*$}, {$D_{s1}$} and {$D_{s2}^*$} are given in Sec. \ref{sec:results} with a summary and conclusion in Sec. \ref{sec:conclusions}. Some details of the interpolating operators are discussed in the Appendix.

\section{Analysis tools}\label{sec:tools}

\subsection{Simulation parameters}

Two different methods (distillation  \cite{Peardon:2009gh} and stochastic distillation \cite{Morningstar:2011ka}) are employed on two different ensembles of gauge configurations. The parameters of the ensembles are given in Table \ref{tab:ensembles}.

\begin{table}[t]
\begin{ruledtabular}
\begin{tabular}{cll}
			&	Ensemble (1)	&	Ensemble (2)            \cr
\hline			
$N_L^3\times N_T$ & $16^3\times32$&	$32^3\times64$	\cr
$N_f$ 			&	2		&	2+1				\cr
$a$[fm] 		&	0.1239(13)	&	0.0907(13)		\cr
$L$[fm] 		&	1.98(2)		&	2.90(4)				\cr
$L m_\pi$		&	2.68(3)	&	2.29(10)			\cr
\#configs 		&	279		&	196				\cr
$a m_\pi$		&	0.1673(16)	&	0.0717(32)		\cr
$a m_K$			&	0.3467(8)	&	0.2317(6)			\cr
\hline			
$\kappa_u$(dyn)	&	0.12830		&	0.13781 			\cr
$\kappa_u$(val)	&	0.12830		&	0.13781 			\cr
$c_{sw}$			&	1.00000		&	1.71500 			\cr
\hline			
$\kappa_s$(dyn)	&	--			&	0.13640				\cr
$\kappa_s$(val)	&	0.12610		&	0.13666			\cr
$c_{sw}$			&	1.00000		&	1.71500			\cr
\hline			
$\kappa_c$(val)	&	0.12300		&	0.12686			\cr
$c_{sw}$			&	1.75218		&	1.64978			\cr
	
\end{tabular}
\end{ruledtabular}
\caption{\label{tab:ensembles} The gauge configurations of ensemble (1) have been produced by \cite{Hasenfratz:2008fg,Hasenfratz:2008ce} (for more details see \cite{Lang:2011mn}), Those of ensemble (2) are due to the PACS-CS collaboration \cite{Aoki:2008sm}. In the table $N_L$ and $N_T$ denote the number of lattice points in spatial and time directions, $N_f$  the number of dynamical flavors and $a$ the lattice spacing. The Pion mass for ensemble (2) is taken from \cite{Aoki:2008sm}.}
\end{table}

\noindent{\bf Ensemble  (1)} has $N_f=2$ dynamical light quarks, a Pion mass of 266 MeV and a coarser lattice spacing. It uses improved Wilson fermions and had been produced in a reweighting study \cite{Hasenfratz:2008fg,Hasenfratz:2008ce}.  The lattice size $16^3\times 32$ and physical volume are small enough that we can use the standard distillation method  \cite{Peardon:2009gh} with a complete set of perambulators (one for each time slice source vector set). We have used this set previously and refer the readers to these publications \cite{Lang:2011mn,Lang:2012sv,Mohler:2012na} for further details.
The gauge links are four-dimensional normalized hypercubic (nHYP) smeared \cite{Hasenfratz:2007rf} with the same parameters used for generating the gauge configurations ($(\alpha_1,\alpha_2,\alpha_3)=(0.75, 0.6, 0.3)$). For the calculation of the eigenmodes and the interpolating fields containing covariant derivatives, we used no additional link smearing. 

\noindent{\bf Ensemble (2)} with $N_f=2+1$ dynamical quarks has been generated by the PACS-CS collaboration \cite{Aoki:2008sm}. Sea and valence quarks are non-perturbatively improved Wilson fermions. It has finer lattice spacing and a Pion mass of 156 MeV. Due to the large lattices size $32^3\times 64$ and larger physical volume we used stochastic distillation \cite{Morningstar:2011ka}. For the calculation of the eigenmodes and the interpolating fields containing covariant derivatives, we used 3D hypercubic smearing (HYP)  \cite{Hasenfratz:2001hp,Hasenfratz:2001tw}  in each time slice.
 
For ensemble (1) the determination of the lattice spacing was discussed previously \cite{Lang:2011mn}. For ensemble (2) the value $a = 0.0907(13)$ fm determined by the PACS-CS collaboration \cite{Aoki:2008sm} is used. In the tables we give the  systematic errors due to the definition of the scale  based on those given in Table \ref{tab:ensembles}.

A word of caution is in order about the determination of the lattice scale on
both ensembles. While we used above determinations for all values quoted in
this paper, we compared those values of the lattice spacing with the ones we
obtain calculating $w_0$ from the Wilson gradient flow method  (for the method, c.f.,
\cite{Luscher:2010iy,Borsanyi:2012zs}) and taking suitable literature
values for the physical value of $w_0$ from other lattice collaborations. Taking the physical 2 flavor value from
the Alpha collaboration \cite{Bruno:2013gha} and an estimate of the quark mass dependence
of $w_0$ from the Budapest-Marseille-Wuppertal (BMW) collaboration (Equation (6.1) in
\cite{Borsanyi:2012zs}) we obtain a 2.6\% smaller lattice spacing for ensemble (1). Assuming the same
quark mass dependence and the physical 2+1 flavor values from BMW \cite{Borsanyi:2012zs} or the 2+1+1
flavor value from HPQCD \cite{Dowdall:2013rya} we end up with lattice spacings $a$ that are
4.4\% or 2.1\% larger than the value determined by PACS-CS for ensemble
(2). We stress that a detailed investigation of scale setting on these
lattices is beyond the scope of our current paper and that the values quoted
in this paragraph should only serve to illustrate that there is a potential
additional uncertainty in setting the scale which we are currently not able to
take into account. Notice that a change in scale
would necessitate a retuning of charm and strange quark hopping parameters which makes
an ad-hoc estimate of the full scale setting uncertainty on final observables
difficult.

\subsubsection{The strange quark mass}

In ensemble (1) the strange quark is included only as a valence quark in the hadron propagators. To determine the strange quark hopping parameter $\kappa_s$ we calculated the connected part of the $\phi$ meson. The tuning has
been discussed in Ref. \cite{Lang:2012sv} and with the final value of $\kappa_s$ we obtain $m_\phi^{lat}=1015.8\pm10.8~$MeV which has to be compared to the experimental mass $m_\phi^{exp}=1019.455\pm0.020~$MeV. 

For ensemble (2) the dynamic strange quark mass used in \cite{Aoki:2008sm} differs significantly from the physical value. We therefore use a partially quenched strange quark $m_{s}^{val}\ne m_{s}^{sea}$ and determine the hopping parameter $\kappa_s^{val}$ by minimizing the difference of the $\phi$ meson mass from the experimental mass and the difference of the unphysical $\eta_s$ meson from the value expected from a high-precision lattice determination \cite{Dowdall:2013rya} $m_{\eta_s}=688.5(2.2)$. The determinations agree excellently and yield the value for $\kappa_s$ in Table \ref{tab:ensembles}. The mass of the $\phi$ and $\eta_s$ mesons for this value of $\kappa_s$ are listed along with a number of mass splittings in Table \ref{deltamtable}.

\subsubsection{The charm quark mass}\label{sub:cmass}

The charm quark is treated as valence quark in both ensembles. The Fermilab method \cite{ElKhadra:1996mp,Oktay:2008ex} is used in an approach similar to \cite{Burch:2009az,Bernard:2010fr}. Details of the approach used along with results for ensemble (1) have been published previously in \cite{Mohler:2012na} and we refer the reader to this publication for information on the method. Within this approach mass splittings in the $D_s$ spectrum are expected to be close to physical and one therefore compares values of $m-\bar{m}$ to experiment. Here $\bar m=\tfrac{1}{4}(m_{D_s}+3m_{D_s^*})$ is the spin-averaged ground state mass.

\begin{table}[b]
\begin{center}
\begin{ruledtabular}
\begin{tabular}{c|cc}
\T\B & Method (1) & Method (2)\\
\hline
\T\B $M_1$ & 1.20438(15) & 1.20436(15)\\
\T\B $M_2$ & 1.4073(59) & -- \\
\T\B $M_4$ & 1.270(63) & -- \\
\T\B $\frac{M_2}{M_1}$ & 1.1685(49) & 1.1632(42)\\
\hline
\T\B $M_2 [GeV]$ & 3.062(13)(44) & 3.048(11)(44)\\
\hline
\T\B Exp $[GeV]$ & \multicolumn{2}{c}{$3.06861(18)$}\\
\end{tabular}
\end{ruledtabular}
\end{center}
\caption{\label{charm_tuning}Fit parameters obtained for spin-averaged charmonium (ensemble (2)) with both tuning methods from \cite{Mohler:2012na}. The values in the last two rows are in GeV, while all other values are in lattice units. The first error on the kinetic mass $M_2$ is statistical while the second error is from the scale setting. The results for $M_4$ are not used in our setup. The last row contains the experimental value from \cite{Beringer:1900zz}.}
\end{table}

\begin{table}[bht]
\begin{center}
\begin{ruledtabular}
\begin{tabular}{c|cc}
 \T\B & Method (1) & Method (2)\\
\hline
\T\B $M_1$ & 0.84606(28) & 0.84601(28)\\
\T\B $M_2$ & 0.9336(105)& -- \\
\T\B $M_4$ & 0.959(71) & -- \\
\T\B $\frac{M_2}{M_1}$ & 1.1035(122) & 1.0978(101)\\
\hline
\T\B $M_2 [GeV]$ & 2.031(23)(39) & 2.021(19)(29)\\
\hline
\T\B Exp $[GeV]$ & \multicolumn{2}{c}{$2.07635(38)$}\\
\end{tabular}
\end{ruledtabular}
\end{center}
\caption{\label{hs_tuning}Same as Table \ref{charm_tuning} but for charm-strange ($D_s$) mesons.}
\end{table}

\begin{table}[bht]
\begin{center}
\begin{ruledtabular}
\begin{tabular}{c|cc}
 \T\B & Method (1) & Method (2)\\
\hline
\T\B $M_1$ & 0.80466(137) & 0.80469(138)\\
\T\B $M_2$ & 0.884(50)& -- \\
\T\B $M_4$ & 0.98(38) & -- \\
\T\B $\frac{M_2}{M_1}$ & 1.099(61)& 1.099(55) \\
\hline
\T\B $M_2 [GeV]$ & 1.923(108)(28) & 1.924(97)(28)\\
\hline
\T\B Exp $[GeV]$ & \multicolumn{2}{c}{$1.97512(12)$}\\
\end{tabular}
\end{ruledtabular}
\end{center}
\caption{\label{hl_tuning}Same as Table \ref{charm_tuning} but for charm-light ($D$) mesons. Notice that the value for $M_2$ in physical units is based on a heavier than physical light-quark mass.}
\end{table}

In the simplified form that we use \cite{Burch:2009az,Bernard:2010fr}, only the charm quark hopping parameter $\kappa_c$ is tuned non-perturbatively, while the clover coefficients $c_E$ and $c_B$ are set to the tadpole improved value $c_E=c_B=c_{sw}^{(h)}=1/{u_0^3}$, where $u_0$ denotes the average link. There are several ways of setting $u_0$  and we opt to use the Landau link on unsmeared gauge configurations.

To tune the hopping parameter $\kappa_c$, the spin-averaged kinetic mass ($M_2$ below) of either heavy-light mesons or charmonium is tuned to be close to the value obtained in experiment. To disentangle the tuning procedures for charmonium from the tuning of the strange quark mass described above, we use the spin average of the 1S charmonium states and therefore tune $(m_{\eta_c}+3m_{J/\Psi})/4$ to its physical value. Determining the charm quark hopping parameter therefore translates into determining the kinetic mass $M_2$ from the lattice dispersion relation \cite{Bernard:2010fr}
\begin{align}
E(p)&=M_1+\frac{\mathbf{p}^2}{2M_2}-\frac{a^3W_4}{6}\sum_ip_i^4-\frac{(\mathbf{p}^2)^2}{8M_4^3}+ \dots\;,
\label{disp}
\end{align} 
where $\mathbf{p}=\frac{2\pi}{L}\mathbf{q}$ for a given spatial extent $L$. 

In \cite{Mohler:2012na} two methods for fitting to the data are used and in the following we present the values obtained from methods (1) and (2) of \cite{Mohler:2012na} on ensemble (2) for our final choice of  $\kappa_c$ listed in Table \ref{tab:ensembles}. Unlike the corresponding values for ensemble (1) found in Tables II, III and IV of \cite{Mohler:2012na}, the results presented here take into account the correlation between energy values at different momentum.  For our final data we use method (1) where the coefficient $W_4$ of the term breaking the rotational symmetry is neglected. By comparison with method (2) we find that it is negligibly small and we stress that the results from both methods are consistent within uncertainties for both ensembles. For the corresponding spin averages for ensemble (2) we use the values of $M_1$ in the tables.

It is worth noting that on both ensembles a physical charmonium mass leads to somewhat lighter than physical heavy-light and heavy-strange meson masses. This is a result of subleading discretization effects which differ between charmonium and heavy-light states. Therefore we stress that our results will not be precision results, which would need a continuum extrapolation.

\begin{table}[t]
\begin{ruledtabular}
\begin{tabular}{clll}
			&	Ensemble (1)	&	Ensemble (2)     & Experiment       \cr
\hline
$m_\pi$ 		&	266(3)(3)	&	156(7)(2)	& 139.5702(4)		\cr
$m_K$		&	552(1)(6)	&	504(1)(7)	& 493.677(16)	\cr
$m_\phi$		&      1015.8(1.8)(10.7)	&      1018.4(2.8)(14.6)& 1019.455(20)	\cr
$m_{\eta_s}$	&	732.3(0.9)(7.7)        &       692.9(0.5)(9.9)	& 688.5(2.2){\bf *}	\cr
\hline	
$m_{J/\Psi}-m_{\eta_c}$ & 107.9(0.3)(1.1) & 107.1(0.2)(1.5)& 113.2(0.7)\cr
$m_{D_s^*}-m_{D_s}$   & 120.4(0.6)(1.3) & 142.1(0.7)(2.0) & 143.8(0.4)\cr	
$m_{D^*}-m_{D}$      & 129.4(1.8)(1.4)& 148.4(5.2)(2.1) & 140.66(10)\cr
$2m_{\overline{D}}-m_{\overline{\bar{c}c}}$ & 890.9(3.3)(9.3)& 882.0(6.5)(12.6) & 882.4(0.3)\cr
$2M_{\overline{D_s}}-m_{\overline{\bar{c}c}}$& 1065.5(1.4)(11.2) & 1060.7(1.1)(15.2)& 1084.8(0.6)\cr
$m_{D_s}-m_{D}$ & 96.6(0.9)(1.0) & 94.0(4.6)(1.3) & 98.87(29)\cr
\end{tabular}
\end{ruledtabular}
\caption{\label{deltamtable}Various meson masses and mass splittings (in MeV) compared to their physical values from \cite{Beringer:1900zz}. For the Pion and Kaon we compare to the charged mesons. For the unphysical $\eta_s$ our values are compared to the value from HPQCD \cite{Dowdall:2013rya} at the physical point (denoted by the asterisk). The error bars indicate the uncertainty due to statistics and due to scale setting. The results do not include infinite volume or continuum extrapolations and are therefore not precision results, but demonstrate a qualitative agreement with experiment.}
\end{table}

 To check our strange and charm quark mass we list further relevant observables in Table \ref{deltamtable}. Note that for these numbers only, the spin averages are not from the dispersion relation fits but instead are the ones derived from correlators at momentum zero. These two choices agree well for all values presented. For ensemble (2) mass differences involving mesons with one or more charm quarks are all close to their respective experiment values, however for ensemble (1) (coarser lattice spacing, containing only two flavors of light dynamical quarks), the $D$ and $D_s$ hyperfine splittings deviate substantially from the experiment value.

\subsection{Dispersion relation}\label{sec:disp_relations}

For the analysis of the phase shifts discussed in Sections \ref{sec:phase_shift} and \ref{sec:eff_range} the dispersion relations for the Kaon ($K$) and heavy meson ($M$) are needed. They are given by
 \be\label{eq:dr_for_D}
E_M(p)= M_1+\frac{{\mathbf{p}}^2}{2M_2}-\frac{{(\mathbf{p}}^2)^2}{8M_4^3}\FC
\ee
\be\label{eq:dr_for_K}
E_K(p)=\sqrt{m_K^2+\mathbf{p}^2}\FC
\ee
which corresponds to the dispersion relation already used for the heavy meson in method (1) of our tuning procedure in \ref{sub:cmass}. While Table \ref{hl_tuning} lists the value obtained for the spin average, we also need the values for $D$ and $D^*$ mesons separately, and they are listed in Table \ref{dispresults}.

\begin{table}[t]
\begin{ruledtabular}
\begin{tabular}{cll}
			&	Ensemble (1)	&	Ensemble (2)            \cr
\hline			
$D: a M_1$		&	0.9801(10)		&	0.7534(12)		\cr
$D: a M_2$		&	1.107(12)		&	0.828(39)		\cr
$D: a M_4$		&	1.107(27)		&	0.89(23)		\cr
\hline			
$D^*: a M_1$		&	1.0629(13)	&	0.8217(16)		\cr
$D^*: a M_2$		&	1.267(21)		&	0.905(66)		\cr
$D^*: a M_4$		&	1.325(68)		&       0.98(51)		\cr
\end{tabular}
\end{ruledtabular}
\caption{\label{dispresults}The parameters for the dispersion relation \eq{eq:dr_for_D} for $D$ and $D^*$ for both ensembles.}
\end{table}

We obtained the energy values from correlators at various momenta $0\le |a\, p|\le 2\sqrt{5} \,\pi/N_L$. For the
vector meson one has to take care of the possible irreducible representations (irreps) of the
symmetry groups for the moving frame \cite{Leskovec:2012gb}.

\subsection{Distillation and stochastic distillation}

In this section our notation for the distillation \cite{Peardon:2009gh} and the stochastic distillation approach \cite{Morningstar:2011ka} is presented. For stochastic methods see also Ref. \cite{Foley:2005ac}.

\subsubsection{Distillation method}

The basic idea is to use for the quark sources the eigenvectors of the spatial lattice Laplacian in each time slice.
We denote an eigenvector in the time slice $t$ by $v_i(\vec{x},c; t)$ ($i$ denotes the index of the Laplacian eigenvector, $t$ denotes the time slice $0\ldots N_T-1$, $\vec{x}$ denotes the spatial lattice position, while $c$ denotes the color index $1\ldots n_c=3$). 

We arrange all eigenvectors in a matrix $V(t)$ with the eigenvectors as $ n_c \,N_L^3$ columns.
The unit operator may be written in terms of its spectral decomposition through the eigenvectors,
\be\label{eq:unit_op}
V V^\dagger=\unitmatrix 
\ee
or, explicitly
\be
v_i(\vec{x},c;t)\,v^{*}_i(\vec x',c';t)=
 \delta_{\vec{x}\vec{x}'}\,\delta_{cc'}\FC
\ee
where we sum over paired indices. The sum over all eigenvectors is truncated to a subset $n_v \ll n_c \,N_L^3$ and instead of the delta function $\delta_{\vec x,\vec x'}$ one obtains a Gaussian-like shape \cite{Peardon:2009gh}. 

We define the standard perambulators.
\begin{align}\label{peram1}
\tau_{ij}^{{\overline\alpha}{\overline\beta}}(t',t)&=v^{*}_i(\vec{x}',c'; t') \, G^{{\overline\alpha} {\overline\beta}}(\vec{x}',c',t';\,\vec{x}, c,t)\,
 v_j(\vec{x},c; t)\nonumber \\
&= v^*_i(\vec{x}',c',{\overline\alpha}; t') u^{{\overline\alpha}}_{\alpha'} \times\nonumber\\
&\quad\ G^{\alpha' \beta'}(\vec{x}',c',t';\,\vec{x}, c,t)\,
 v_j(\vec{x},c,{\overline\beta}; t)  u^{{\overline\beta}}_{\beta'}\FD
\end{align}
Here $G$ is the usual quark propagator and in the second step we have introduced unit length spinors $u^{(1)}=(1,0,0,0)$,
$u^{(2)}=(0,1,0,0)$ etc. which makes the role of the spin indices explicit and which facilitates the later discussion of stochastic distillation. In this expression and also further down the notation 
${\overline\alpha}$ indicates that in this case the index 
is considered fixed and not summed over. The extra index in the vector is trivial, $v_j(\vec{x},c,\beta; t)\equiv v_j(\vec{x},c; t)$.
The perambulators are thus propagators between quark sources $v_j(\vec{x},c; t)$ and $v^{*}_i(\vec{x}',c'; t')$.

Once one has determined the perambulators $\tau$, the hadron propagator can be evaluated with high flexibility in the interpolators. Projection to spatial momenta, different Dirac and color structure and derivatives all can be defined independent of the perambulators.

Consider, e.g., meson interpolators of the form
\be
M(\vec p,t)= \overline u^{\alpha}(\vec x,a,t) \,\Gamma_{a b}^{\alpha  \beta}(\vec x, \vec y;\vec p,t) \,
d^{\beta}(\vec y,b,t)\FC
\ee
where summation over $\vec x$, $\vec y$ and pairs  of colors $(a,b)$ and Dirac indices ($\alpha,\beta$) is implied. The meson kernel
includes projection to spatial momentum $\vec p$ as well as possible derivatives, color  and Dirac structures.
We omit all indices for short-hand notation, writing
\be
M(\vec p,t)= \overline u \,\Gamma\, d\FD
\ee
Distillation introduces the approximate unit operator (quasi smearing operator) \eq{eq:unit_op} in the form
\begin{align}
\label{eq:meson_kernel_brief}
M(\vec p,t)= &\overline u\, V V^\dagger\, \Gamma\, V V^\dagger\, d\FD
\end{align}
Propagators for such interpolators may then be written 
\begin{align}\label{eq:corr_mat_dist}
&\langle M(\vec p,t') \,M^\dagger(\vec p,t)\rangle \nonumber\\
&=
\langle 
 \overline u \,V V^\dagger \,\Gamma\, V V^\dagger \,d
  \overline d \,V V^\dagger\, \Gamma^\dagger \,V V^\dagger\, u
   \rangle \nonumber\\
   &=
-\langle 
( V^\dagger \Gamma V)\,
 ( V^\dagger \,d  \overline d\, V )\,
 (V^\dagger \Gamma^\dagger V)\,
 ( V^\dagger \,u  \overline u\, V) \rangle\nonumber\\
 &= 
- \trace \left[
( V^\dagger \,\Gamma\, V)
( V^\dagger \,G_d \,V )
(V^\dagger \,\Gamma^\dagger\, V)
( V^\dagger \,G_u \,V)\right] \nonumber\\
 &= 
-\trace \left[\phi(t') \tau(t',t) \phi(t) \tau(t,t')\right]\FD
\end{align}
The brackets $\langle\ldots\rangle$ denote the integration over the Grassmann variables $u, \overline u, d, \overline d$ and the extra minus sign
is due to anti-commuting $\overline u$ from left to right. The time slice positions have been indicated for convenience.
We have introduced the meson kernel $\phi$ for a given time slice denoted by
\begin{align}\label{eq:def_phi}
\phi &= V ^\dagger\Gamma V  \FC\textrm{~or~}\\
\phi^{\alpha \beta}_{ij} &= v^*_i(\vec x,c)\Gamma^{\alpha \beta}(\vec x,c;\,\vec x',c') 
v_j(\vec x',c') \FC\nonumber
\end{align}
where $V$ and $\Gamma$ also live on that time slice.

Using $\gamma_5$-hermiticity, we have
\be
\tau(t,t')=\gamma_5\tau(t',t)^\dagger\gamma_5\textrm{~~or, short~~}
\tau_{ij}=\gamma_5\tau_{ji}^*\gamma_5\FD
\ee

\subsubsection{Stochastic distillation}

In distillation the number of Laplacian eigenvectors $n_v$ grows with the physical volume in order to keep the source profile constant in physical size.\footnote{As a rule of thumb one needs more than $\mathcal O(64)$ vectors for a box with spatial size $~2$ fm. For higher momenta even more vectors are needed and eventually the approach may become inefficient.} This leads to technical problems for large volumes. As a remedy to this a stochastic version of distillation was suggested in Ref. \cite{Morningstar:2011ka}. The number of sources $n_v$ is reduced by using stochastic combinations instead of the eigenvectors. We discuss here our implementation of that formalism.

For the notation we now define (for each quark species) on each time slice vectors $\rho^{[r]}$ of $4 n_v$ random numbers,
\be
\rho^{\alpha [r]}_i\equiv 
 \quad\textrm{with}\quad \left[\rho^{\alpha [r]}_i\right]_r=0\FC\quad  \left[\rho^{\alpha [r]}_i \rho^{\beta [r]*}_j \right]_r=\delta_{ij} \delta_{\alpha \beta}\FC
\ee
(the greek indices are Dirac indices). We have introduced the average $[\ldots]_r$ over the space $S$ of random numbers $\rho^{[r]}$.
In practice one has $n_r\ll n_v$. For $S$ we use the space of uniformly distributed unimodular complex numbers. The
products $v\cdot \rho^{[r]}$ provide stochastic sources for each $r$.

It is advantageous to partition the source vectors into disjoint parts (indexed
by  $b$). For the projectors $P^{(b)}$ (with $P=P^2$) we use $n_b$ diagonal $n_v\times n_v$ matrices with diagonal elements assuming values 1 or 0, and 
\be
\left[P^{(b)}\right]_b\equiv\sum_{b=1}^{n_b}  P^{(b)}=\unitmatrix_{n_v\times n_v}\FD
\ee
With their help we introduce the rectangular $n_v\times n_b$ matrices
$\eta^{\alpha[r]}$
with the matrix elements
\be
(\eta^{\alpha [r]})_{ib} = \sum_j P^{(b)}_{ij}\rho^{\alpha [r]}_j \FD
\ee
Obviously
\be\label{eq:eta_unity}
[ \eta^{\alpha[r]} \eta^{\beta[r]\dagger}]_{r}=\unitmatrix_{n_v\times n_v}\delta_{\alpha\beta}\FD
\ee
We can write the $4 n_b$ stochastic sources as scalar product
\be
S^{{\overline\alpha} [r]}_{b}(\vec{x},c;t) u^{{\overline\alpha}}_\beta=\sum_i  v_i(\vec{x},c;t) \;\eta^{{\overline\alpha} [r]}_{ib}\;u^{{\overline\alpha}}_\beta \FC
\ee
and replace $v_i(\vec{x},c,\alpha;t)$ by $S^{{\overline\alpha} [r]}_{b}(\vec{x},c;t)$ in \eq{peram1}. In this expression and also further down the notation ${\overline\alpha}$ indicates that in this case the index is considered fixed and not summed over.
It will be shown below that Wick contractions expressed in terms of sources $S$ will reduce to the expressions in full distillation after averaging over noises $r$.
One now introduces stochastic perambulators\footnote{The so defined perambulators
are ``half''-stochastic; one could also define them symmetrically.}
\begin{align}
T^{{\overline\alpha}{\overline\beta} [r]}_{ib}(t,t') &= 
v^*_i(\vec{x},c,{\overline\alpha};t)u^{{\overline\alpha}}_{\alpha'}\times\\
& G^{\alpha' \beta'}(\vec{x},c, t; \vec{x}',c', t') S^{{\overline\beta} [r]}_{b}(\vec{x}',c';t') u^{\overline\beta}_{\beta'}\FC\nonumber
\end{align}
where the noise vectors live in the corresponding time slices. We could recover the standard
perambulators through
\be
\label{Tandtau}
\sum_b \left[T^{\alpha\beta [r]}_{ib}(t,t')\eta^{*\gamma[r]}_{jb}\right]_{r}=\tau^{\alpha \gamma}_{ij}(t,t')\FD
\ee
The stochastic perambulators are propagators from the $(n_r n_b)$ stochastic source vectors  $S^{[r]}_b$ to the sink vectors $v_i$. We will express all hadron propagators in terms of $T$.

In our approach we use two types of stochastic sources. The first type (A) locates the sources on just one
time slice.  For the partitioning projectors (eigenvector interlacing) in a given time slice we choose
\be
P^{(b)}_{nm}(t)=\delta_{nm} \sum_{k=0}^{n_i-1} \delta_{b+k n_b,m}\FC
\ee
where $b$ runs from 1 to $n_b=n_v/n_i$ and $n_i$ is the number of non-vanishing entries in each $P^{(b)}$.
For each configuration we calculate $N_T/n_{ti}$ perambulators for the time slices with distance $n_{ti}$ located
at $t=0$, $n_{ti}$, $2n_{ti}$, etc.

The second set of sources (B) are time-interlaced sources and have support simultaneously on several time slices with distance $n_{ti}$. There are $k=0\ldots (n_{ti}-1)$ such sources where the $k$-th of those has support on  $t=k,\, k+n_{ti}$, $k+2n_{ti}, \ldots$. The perambulators of type (B) are used for the backtracking quark lines on the sink time slices.
For the time-interlacing partitioning projectors $P^{(b)}$ we thus have
\be
P^{(k)}_{nm}= \sum_{\delta=0}^{N_T/n_{ti}-1} P^{(b)}_{nm}(k+\delta \,n_{ti})\FD
\ee

Consider the correlation matrix from Eq. \eq{eq:corr_mat_dist}, replacing $\tau$ by the stochastic perambulator $T$ via (\ref{Tandtau}) 
\begin{align}
&\trace \left[\phi(t') \tau(t',t) \phi(t) \gamma_5\tau(t',t)^\dagger\gamma_5 \right]\\
&=
\trace \left[\phi(t')\left [T^{[r]}(t',t)\eta^{[r]\dagger}\right]_{r} \phi(t) \gamma_5 \left[T^{[r']}(t',t)\eta^{[r']\dagger}\right]_{r'}^\dagger\gamma_5\right]\FD\nonumber
\end{align}
For each $r$, $r'$ this may be rearranged
\begin{align}\label{eq:stoch_peramb_corr}
&\trace \left[\phi(t') T^{[r]}(t',t)  \left(\eta^{[r]\dagger} \phi(t)  \eta^{[r']}\right)\gamma_5 T^{[r']\dagger}(t',t) \gamma_5\right]_{r,r'}\nonumber\\
&=
\trace \left[\phi(t') T^{[r]}(t',t) \widehat \phi^{[r,r']}(t) \gamma_5 T^{[r']\dagger}(t',t) \gamma_5\right]_{r,r'}\FC
\end{align}
where we have introduced a modified meson kernel operator
\begin{align}\label{eq:def_widehatphi}
\widehat \phi^{\overline\alpha{\overline\beta}[r,r']}_{bb'}(t)&=
S^{{\overline\alpha} [r]*}_{b}(\vec{x},c;t) u^{{\overline\alpha}}_{\alpha'}
\Gamma^{\alpha'\beta'}(\vec x,c;\,\vec x',c') \times\nonumber\\
&\qquad\qquad S^{{\overline\beta} [r']}_{b'}(\vec x ',c';t) u^{{\overline\beta}}_{\beta'}
\end{align}
at the source time slice ($\overline\alpha, \overline\beta$ are external indices not summed).
 
An alternative prescription to arrive at this form is to insert \eq{eq:eta_unity} into the meson interpolator \eq{eq:meson_kernel_brief} at the source, giving
 \be
 M(\vec p,t)=[ \overline u \,V  \eta^{[r]} \eta^{[r]\dagger} V^\dagger \,\Gamma\, V \eta^{[r']} \eta^{[r']\dagger} V^\dagger \,d]_{r,r'}\FD
 \ee
This then together with \eq{eq:corr_mat_dist} gives \eq{eq:stoch_peramb_corr}. This way the ``smeared'' quark $VV^\dagger q$ has been replaced by
$V\eta\eta^\dagger V^\dagger q$ at the source.

Since it is important for the practical implementation, let us summarize the range of indices of the
terms.
\begin{itemize}
\item
$T^{\alpha\alpha'[r]}_{jb}(t',t)$ for each $b$ and $r$ has a ``left'' index, $j$, running over $1\ldots  n_v$ and a ``right'' index , $b$, running over $1\ldots  n_b$; it also has left and right Dirac indices $\alpha$ and $\alpha'$ (inherited from $\tau$), 
\item
For $\phi_{ij}^{\alpha\alpha'}(t)$ both indices $i,j$ run over $1\ldots n_v$ and the Dirac indices over $1\ldots 4$.
\item
For $\widehat \phi^{\alpha\alpha'[r,r']}_{bb'}(t)$ the  indices $b, b'$  run over $1\ldots  n_b$ and the Dirac indices over $1\ldots 4$. 
\item
For diagrams with backtracking quark lines we also need another (rectangular) version of the meson kernel:
$\overline\phi^{\alpha\alpha'[r]}_{bi}(t)$,where $b$ run over $1\ldots n_b$, the other  over $1\ldots n_v$ and the Dirac indices over $1\ldots 4$.
\end{itemize}

In our implementation on the PACS-CS ensemble of lattices of size $32^3\times 64$ we use $n_v=192$, $n_i=16$ (thus $n_b=12$) and 
for the time interlacing $n_{ti}=8$. For each gauge configuration we therefore compute 8 stochastic perambulators of type (A)
for the time slices $0, 8, 16,\ldots,56$ and 8 time-interlaced perambulators (B) with simultaneous support on 8 time slices each, as discussed above.

For each quark species we have $n_r=4$ random vectors. One has to use different vectors for the different quarks lines in a diagram. We average over permutations of the stochastic perambulators for different $r$.
In total the Dirac operator has to be inverted $2 n_b (N_T/n_{ti}) n_D n_r=3072$ times for each quark species. For this we use the highly efficient SAP-GCR inverter from L\"uscher's DD-HMC package \cite{Luscher:2007es,Luscher:2007se}.

For the calculation of the eigenmodes we use the PRIMME package \cite{Stathopoulos:2009PPI}. In particular, the routine \verb+JDQMR_ETOL+ results in a fast determination for a small to moderate number  of eigenmodes. For a larger number of eigenmodes the Arnoldi/Lanczos method \cite{Lehoucq:1998xx} (and variants) eventually outperforms this method. For the methods implemented in PRIMME we also tried a preconditioner using Chebychev polynomials, very similar to the method described in \cite{Morningstar:2011ka}. The preconditioner greatly improved the performance of the Arnoldi implementation in  PRIMME while some other methods were largely unaffected. For ensemble (1) we used \verb+JDQMR_ETOL+ without preconditioner while we used Arnoldi with preconditioner for ensemble (2).

\subsubsection{Sample diagram}

\begin{figure}[t]
\begin{center}
\includegraphics*[width=0.35\textwidth,clip]{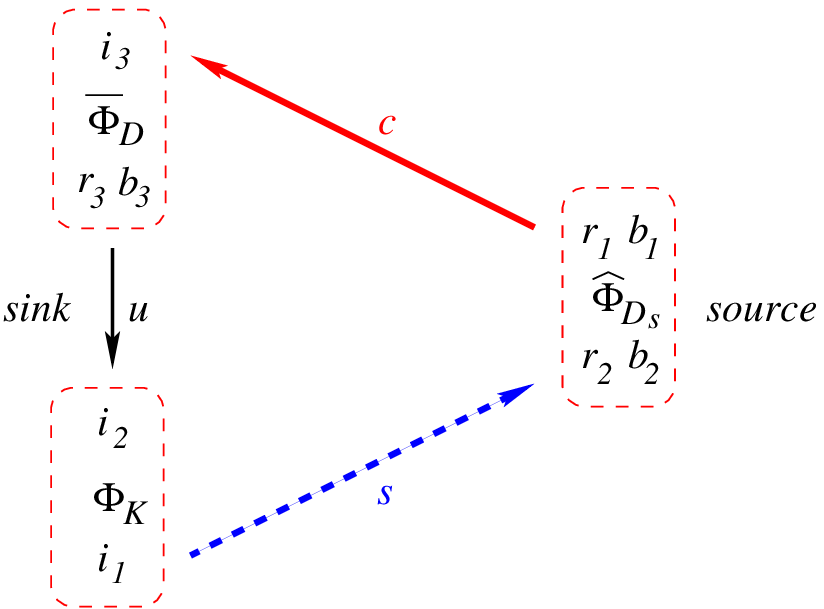}
\end{center}
\caption{Diagrammatic representation of the triangle contribution Eq. \eq{eq:triangle_Stoch} for
$D_s^+\to D^0 K^+$.}
\label{fig:triangle_D_s_DK}
\end{figure}

As an example for a diagram involving backtracking quark lines we consider the triangle diagram $D_s^+\to D^0 K^+$ corresponding to $c\sbar\to c\ubar u\sbar$. The diagram in Fig.~\ref{fig:triangle_D_s_DK} should be read clockwise to be translated to the following expression (we omit the Dirac indices):
\begin{align}\label{eq:triangle_Stoch}
&\trace\left[
(\widehat\phi_{D_s})^{[r_1 r_2]}_{b_1 b_2}(t)
(T_s)^{[r_2]}_{b_2 i_1}(t,t')
(\phi_K)_{i_1 i_2} (t')\right.\nonumber\\
&\qquad\left.
(T_u)^{[r_3]}_{i_2 b_3}(t',t')
(\overline \phi_D)^{[r_3]}_{b_3 i_3}(t')
(T_c)^{[r_1]}_{ i_3 b_1}(t',t) 
\right]\nonumber\\
&=
\trace\left[
(\widehat\phi_{D_s})^{[r_1 r_2]}_{b_1 b_2}(t)
\gamma_5 (T_s)^{[r_2]*}_{i_1 b_2 }(t',t)\gamma_5
(\phi_K)_{i_1 i_2} (t')\right.\nonumber\\
&\qquad\left.(T_u)^{[r_3]}_{i_2 b_3}(t',t')
(\overline \phi_D)^{[r_3]}_{b_3 i_3}(t')
(T_c)^{[r_1]}_{ i_3 b_1}(t',t) 
\right]
\end{align}
In this example the perambulator for the backtracking quark line (at the sink) is of the time-interlaced type (B), the others are of type (A). After the average $[..]_{r_1,r_2,r_3}$ over a large number of random numbers, the expression (\ref{eq:triangle_Stoch}) formally renders the expression in full distillation
\begin{align}
&\trace\left[
(\phi_{D_s})_{i_5 i_6}(t)
(\tau_s)_{i_6 i_1}(t,t')
(\phi_K)_{i_1 i_2} (t')\right.\nonumber\\
&\qquad\left.
(\tau_u)_{i_2 i_3}(t',t')
(\phi_D)_{i_3 i_4}(t')
(\tau_c)_{ i_4 i_5}(t',t)~ \nonumber
\right]~.
\end{align}

\subsection{Evaluation of energy levels}

The discrete energy levels were extracted from correlations between sets of interpolating operators (discussed later) using the  variational method \cite{Michael:1985ne,Luscher:1985dn,Luscher:1990ck,Blossier:2009kd}. For a given quantum channel one measures the Euclidean cross-correlation matrix $C_{ij}(t)=\langle O_i(t)O^\dagger_j(0)\rangle$  between several interpolators living on the corresponding Euclidean time slices. The generalized eigenvalue problem  
\be
\label{eq:gevp}
C(t) v^{(n)}(t)=\lambda^{(n)}(t) C(t_0)v^{(n)}(t)
\ee
disentangles the eigenstates $|n\rangle$. From the exponential decay of the eigenvalues $\lambda_n(t)\sim\exp{(-E_n(t-t_0))}$ one determines the energy values $E_n$ of the eigenstates by exponential fits to the asymptotic behavior. In order to obtain the lowest energy eigenstates and energy levels reliably one needs a sufficiently large set of interpolators with the chosen quantum numbers. 

Formally one expects reliable results for $t$ in a range between $t_0$ and $2 t_0$  \cite{Blossier:2009kd}. 
In practice large values of $t_0$ lead to larger fluctuations and the correlation matrix may not be
positive definite any more. We use values up to $t_0=3$ (the first time slice is at $t=0$) and fit over
a larger range $t_0<t_a\le t \le t_b$ to extract the asymptotic value. In general a 2-exponential fit (one of the exponentials deals with the admixture at small $t$) works over an extended range of $t$ values. We check the reliability of the result by comparing with a 1-exponential fit over a smaller $t$ range (i.e., starting at larger $t_a$). 

For both ensembles $t_0=2$  was sufficient for quantum numbers $0^+$ and $2^+$, while we had to choose $t_0=3$ for $1^+$. We performed correlated fits for all energy levels using either a one or two exponential shape, to make sure our results are not affected by excited state contaminations. The fit ranges and final fit shape chosen are indicated in the tables of results. In the figures in Sect. \ref{sec:results} we show the effective energies
\be
a E^{(n)}_\textrm{eff}(t+\tfrac{1}{2})=\log\frac{\lambda^{(n)}(t)}{\lambda^{(n)}(t+1)}\FD
\ee
The fits, however, are directly to $\lambda^{(n)}(t)$.

All error values come from a single-elimination jack-knife analysis, where the error analysis for $p\cot\delta(p)$ includes also the input from the dispersion relation.
\subsection{Scattering amplitude and phase shift above threshold} \label{sec:phase_shift}
 
Assuming a localized interaction region smaller than the spatial lattice extent L\"uscher has derived a relation \cite{Luscher:1985dn,Luscher:1986pf,Luscher:1990ux,Luscher:1991cf} between the energy spectrum of meson-meson correlators in finite volume and the infinite volume phase shift in the elastic region and in the rest frame,
\be
\label{eq:luescher_z}
\tan \delta(q)=\frac{\pi^{3/2} q}{\mathcal{Z}_{00}(1;q^2)}\FC
\end{equation}
where the generalized zeta function $\mathcal{Z}_{lm}$ is given in \cite{Luscher:1990ux}. The variable  $q$ is defined as the dimensionless product of the momentum  and the spatial lattice size
\be\label{eq:def_of_q}
q=p\frac{L}{2\pi}\FD
\ee
The value of the momentum $p=|\mathbf{p}|$ is obtained from the energy value
\be\label{eq:energy}
E=\sqrt{s}=E_M(p)+E_K(-p)\FD
\ee
where the dispersion relation for $M=D, D^*$ and the Kaon are given in Eqs. \eq{eq:dr_for_D} and \eq{eq:dr_for_K} in Sect. \ref{sec:disp_relations}. We extract the momentum by inverting the dispersion relation.

Eq. \eq{eq:luescher_z} may be written as
\be
p \cot \delta(p)= \frac{2 \mathcal{Z}_{00}(1;(\tfrac{pL}{2\pi})^2)}{L \sqrt\pi}\FC
\ee
which above threshold is the real part of the inverse elastic scattering amplitude $T$.

\subsection{Analytic continuation near threshold} \label{sec:eff_range}

The effective range approximation is a linear (in $p^2$) approximation $(1/a_0 +  r_0 p^2/2)$ of $p \cot \delta(p)$ valid near 
above threshold. The partial wave scattering amplitude $T$ itself has a cusp (in the real part) at threshold. Above threshold we have
\be
T^{-1} \propto p\, \cot \delta(p) - \I\, p\FD
\ee
Below threshold the phase space term $-\I\, p$ becomes real $|p|$, thus the cusp. L\"uscher's formula defines the analytic 
extrapolation of $p \cot \delta(p)$ (see Refs. \cite{Luscher:1985dn,Luscher:1986pf,Luscher:1990ux,Luscher:1991cf,Doring:2011vk}) which is real above and below threshold, that is
\begin{align}
T^{-1}& \propto \frac{2  \mathcal{Z}_{00}}{L \sqrt\pi} - \I p &\textrm{above threshold}&\nonumber\\
T^{-1}& \propto  \frac{2  \mathcal{Z}_{00}}{L \sqrt\pi}+ |p| &\textrm{below threshold}&\FD
\end{align}
Thus the effective range approximation for the quantity $p \cot \delta(p)$ can be continued below threshold as given by the real functions
\be
p \cot \delta(p)= \frac{2  \mathcal{Z}_{00}}{L \sqrt\pi} \approx \frac{1}{a_0} + \frac{1}{2} r_0 p^2+\mathcal{O}(p^4)\FD
\ee
and one can use the two data points (derived from the energy levels above and below threshold) for an approximate determination of its parameters. 

The procedure we employ to obtain the bound state position was proposed by NPLQCD for extracting an $NN$ bound state in future lattice simulations on a single volume \cite{Beane:2010em}. Below threshold $-\I\,p$ becomes $|p|$ or, equivalently, $p\to\I |p|$. In the limit of infinite volume, the $T$-matrix has a pole for real $s$  below threshold when
\be\label{eq:B-condition}
 \I |p_B| \cot\delta(\I |p_B|)+|p_B|=0
\textrm{~~or~}
  \cot\delta(\I |p_B|)=\I\FC
\ee
where $p_B$ is the binding momentum.  It is then determined as a solution of 
\be
\frac{1}{a_0} - \frac{1}{2} r_0 |p_B|^2= -|p_B|\FD
\ee
(See also, e.g., Eq, (4.6) of \cite{Albaladejo:2013aka}).

At finite $L$, the lowest energy level corresponds to
\be\label{eq:B-V-corrections}
\cot \delta(p) = \I +\sum_n \frac{1}{\I |n| |p| L} \E^{-|n| |p| L}\textrm{~~with~~}     n \in N_L^3\FC
\ee
which reproduces \eq{eq:B-condition} for $L\to\infty$. The above relation contains all finite volume corrections $\E^{-|n| |p| L}$  (see for example \cite{Sasaki:2006jn,Beane:2011iw}).

In our simulation the lowest energy levels correspond to values $\cot\delta$ equal to $0.84(2)\I$/$0.86(9)\I$ (for ensembles (1) and (2) in the $J^P=0^+$ channel) and $0.87(1)\I$/$0.88(4)\I$ (for the $1^+$ channel). One way to determine the shift of the bound state position due to finite volume is to simulate several volumes and extrapolate. The second possibility, available on a single volume, is to apply the effective range approximation near threshold.  This allows us to get an estimate of the binding momenta $p_B$ at which the infinite volume pole condition \eq{eq:B-condition} is satisfied; this is preferable compared to simply using the finite volume value of the lower state energy directly. Of course, a future simulation on several volumes would be ideal and would serve as a valuable cross-check.

\subsection{Interpolating operators}

Most lattice studies so far have relied exclusively on $\qbar q$ interpolators. On the other hand we know that the mesons couple to meson-meson channels and the energy spectrum in the quantum channel will be affected at least in the resonance region, in principle everywhere. If a resonance has a small width in the meson-meson channel (i.e., it couples weakly) then the effect will be small and the energy levels will be close to non-interacting ones. This is in particular the case for many heavy quark mesons with small hadronic width and this explains the success of the single hadron interpolator approach. 

We have to stress that in quantum field theory the identification of energy levels with interpolators can be misleading. It is a  combination of interpolating operators used in the simulation that actually defines one physical (eigen)state and its energy level. An example where this is relevant is the $D_{s0}^{*}(2317)$, where without the meson-meson scattering operators the mass obtained from the single hadron approach is too high. Only a detailed analysis, like in Sec. \ref{sec:phase_shift} and Sec. \ref{sec:eff_range} can reveal the physical state.

Depending on the set of interpolators some contributing states may be underrepresented in their weight. Although one expects, that in simulations with fully dynamical quarks the meson-meson intermediate states show up even in $\qbar q$ correlators (of the single hadron approach) most often there is no such signal observed. The addition of meson-meson interpolators for relevant hadronic channels results in a reliable spectrum. Furthermore, the inclusion of scattering operators allows effective study of meson-meson scattering and the emergence of resonances or bound states. This motivates our choice of interpolators listed in App. \ref{app_a}. Our results confirm the importance of scattering channels in a lattice QCD simulation.

We study the $D_s$ channel for the quantum numbers $J^P=0^+$, $1^+$ and $2^+$, the first two near the $DK$ or $D^*K$ thresholds, respectively. For this we use up to eight quark-antiquark interpolating fields and up to three meson-meson interpolators, all projected to total momentum zero.
The interpolating operators, which enter the meson kernels in Eqs. \eq{eq:def_phi} and \eq{eq:def_widehatphi}, are in irreducible representations of the octahedral group $O_h$ and are listed in App. \ref{app_a}. 

\begin{figure}[t]
\begin{center}
\includegraphics*[width=0.4\textwidth,clip]{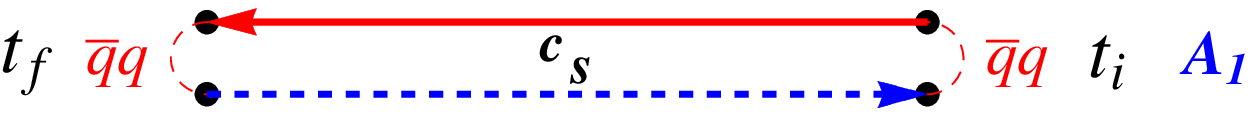}\vspace*{2mm}\\
\includegraphics*[width=0.4\textwidth,clip]{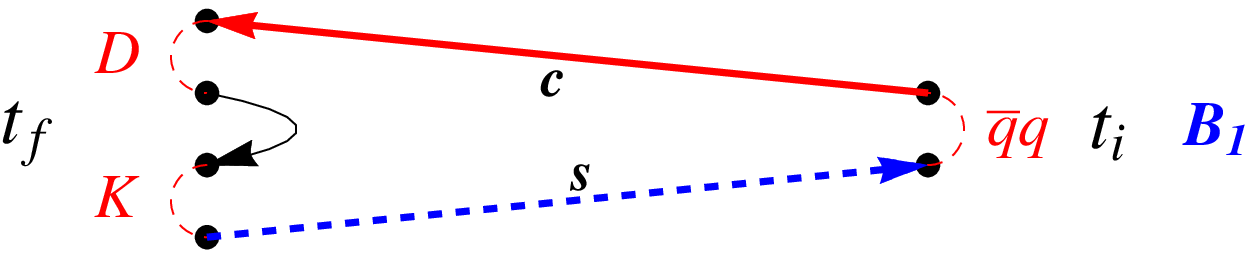}\vspace*{2mm}\\
\includegraphics*[width=0.4\textwidth,clip]{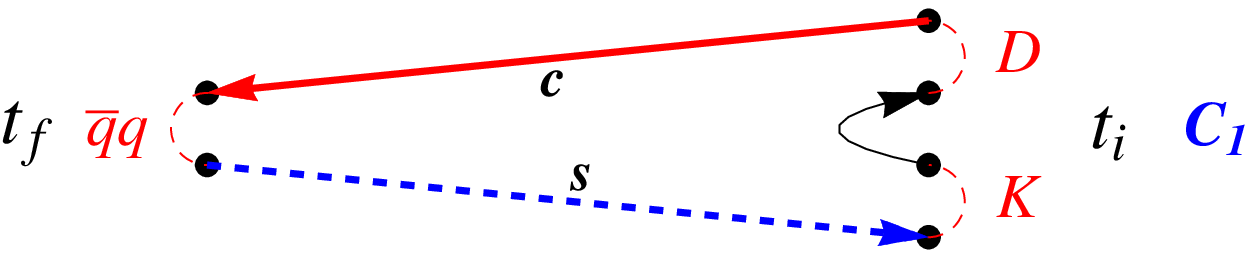}\vspace*{2mm}\\
\includegraphics*[width=0.4\textwidth,clip]{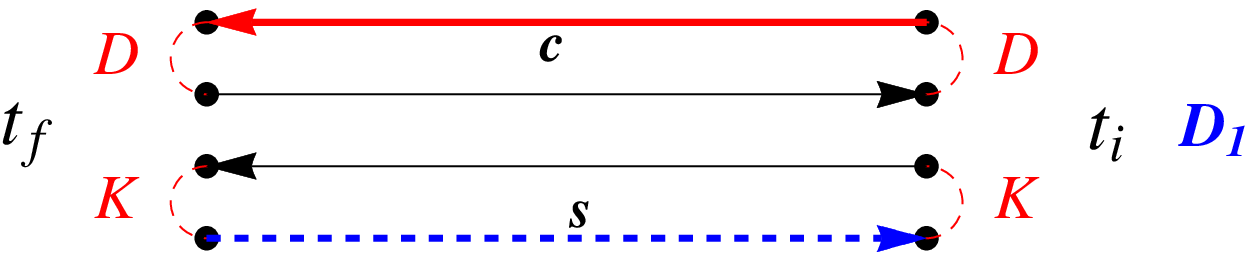}\vspace*{2mm}\\
\includegraphics*[width=0.4\textwidth,clip]{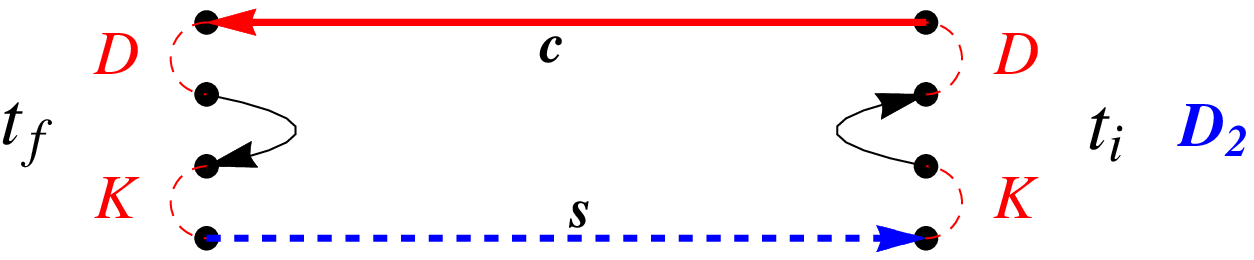}\vspace*{2mm}\\
\end{center}
\caption{Diagrammatic representation of the necessary Wick contractions. Source and sink are indicated by $t_i$ and $t_f$.}
\label{fig:contractions}
\end{figure}

For each spin and parity channel we have a correlation matrix of the form
\begin{equation}
 \bordermatrix{ \qquad         &  \qbar q\textrm{-type}      &  DK\textrm{-type}          \cr
               \qbar q\textrm{-type} &  -A_1            &  -2B_1          \cr
               DK\textrm{-type}      &  -2C_1            &  2\,D_1 - 4\,D_2 \cr}\;,
\end{equation}
which is evaluated using Wick contractions shown symbolically in Fig. \ref{fig:contractions}.

For the $J^P=0^+$ channel, where the $D_{s0}^{*}$ is present, we use four interpolators of type $\qbar q$ and three interpolators of type $DK$ in $s$-wave. These are in the $A_1^+$ irrep and are listed in Table \ref{tab:interpolators} and Eqs. \eq{eq:Op_DK_A1} of App. \ref{app_a}.

In the $J^P=1^+$ channel, where both the $D_{s1}(2460)$ and $D_{s1}(2536)$ are present, we use eight $\qbar q$ interpolators and three $D^*K$ $s$-wave interpolators; all are in the $T_1^+$ irrep and listed in Table \ref{tab:interpolators} and Eqs. \eq{eq:Op_DK_T1} of App. \ref{app_a}. 
 
 The $J^P=2^+$ channel, where the $D_{s2}^*(2573)$ resides, is simulated using only two $\qbar q$ operators in the $T_2^+$ irrep. Interpolators are listed in Table \ref{tab:interpolators} of App. \ref{app_a}.

\section{Results}\label{sec:results}

\subsection{$D_{s0}^*$}

\begin{figure}[t]
\begin{center}
\includegraphics*[height=6.6cm,clip]{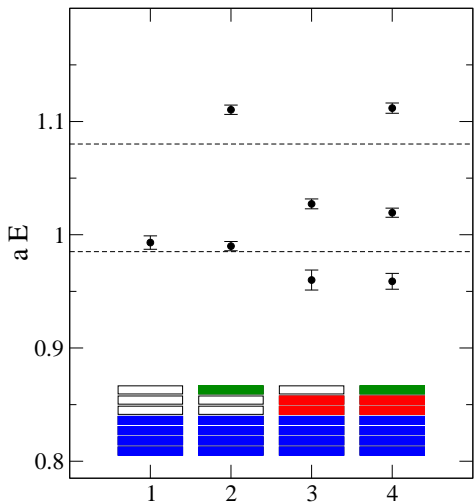}
\end{center}
\caption{$A_1^+ $: Effective energies as obtained for various subsets of operators for ensemble (2). The horizontal broken lines indicate the positions of $D(0)K(0)$ and $D(1)K(-1 )$ in the non-interacting case. The boxes indicate the operators (listed in App. \ref{app_a}) considered in each case (blue: $\qbar q$ , red:  $D(0)K(0)$, 
green: $D(1)K(-1 )$).}
\label{fig:A1_levels_subsets}
\end{figure}

\begin{table*}[tbp]
\begin{center}
\begin{ruledtabular}
\begin{tabular}{c|cccccl|lccccc}
level  & $t_0$ & basis &$\textrm{fit}\atop\textrm{range}$     &$\textrm{fit}\atop\textrm{type}$   & $\tfrac{\chi^2}{d.o.f}$ & $Ea$  & $E-\bar m~$& $(a p)^2$  & $a p ~\cot(\delta)$ &$p^2$       & $ p ~\cot(\delta)$ \vspace{-3pt} \\
 &          &         & &      &                                    &            &       [MeV]      &             &        & [GeV$^2$]      & [GeV]         \\              
\hline                                
\multicolumn{6}{l}{Ensemble (1)}\\
\hline
1 & 2 & $O_{1-7}$ & 4-15 & 2exp$^c$ & 0.07 & 1.2566(28) & 254.1(4.3) & -0.0347(14) & -0.1560(59)  & -0.0881(35) & -0.2484(94) \\
2 & 2 & $O_{1-7}$ & 4-15 & 2exp$^c$ & 0.15 & 1.3922(27) & 470.0(4.0) &  0.0364(14) & -0.1722(74)  &  0.0924(36) & -0.274(12)\\
3 & 2 & $O_{1-7}$ & 4-10 & 2exp$^c$ & 0.17 & 1.6124(69) & 821(11) &  0.1846(52) & -0.526(126)&  0.4682(133)& -0.84(20)\\
\hline                                
\multicolumn{6}{l}{Ensemble (2)}\\
\hline
1 & 2 & $O_{1-7}$      & 3-12 & 2exp$^c$ &  0.44 & 0.9589(70)  & 245(15) & -0.0092 (24) &  -0.082 (19)   &-0.0433 (111)  &-0.178 (41)  \\
2 & 2 & $O_{1-7}$      & 3-11 & 2exp$^c$ &  1.71 & 1.0195(40)  & 377(9)  &  0.0130 (16) &  -0.049 (15)   & 0.0616 (76)   &-0.107 (32)  \\
3 & 2 & $O_{1-7}$      & 3-11 & 2exp$^c$ &  0.66 & 1.1118(45)  & 578(10) &  0.0531 (22) &  -0.053 (49)   & 0.2515 (104)  &-0.114 (106) \\
\end{tabular}
\end{ruledtabular}
\end{center}
\caption{ Energy levels for irrep $A_1^+$. The superscript $c$ indicates a correlated fit and $\bar m=\tfrac{1}{4}(m_{D_s}+3m_{D_s^*})$. \label{tab:A1}}
\end{table*}

\begin{table*}[tbp]
\begin{center}
\begin{ruledtabular}
\begin{tabular}{l|cc|cccc}
set & $a_0^{DK}$ & $r_0^{DK}$ & $(a p_{B})^2$ & $a m_{B}$ & $m_K+m_{D}-m_{B}$  &$m_{B}-\tfrac{1}{4}(m_{D_s}+3 m_{D_s^*}) $ \\
& [fm] & [fm]& & &  [MeV]  & [MeV] \\ 
 \hline                                
\multicolumn{3}{l}{Ensemble (1)}\\
\hline                                
&-0.756(25) & -0.056(31) &-0.0250(17) & 1.2772(32) & 78.9(5.4)(0.8) &287(5)(3)  \\
\hline
\multicolumn{3}{l}{Ensemble (2)}\\
\hline  
 & -1.33(20)&0.27(17)&-0.0060(26)   & 0.9683(76)& 36.6(16.6)(0.5) & 266(17)(4) \\                         
\hline
\multicolumn{3}{l}{Experiment}\\
\hline                                
 &&       &             &                      &  45.1   &  241.5\\
\end{tabular}
\end{ruledtabular}
\end{center}
\caption{\label{tab:A1_EffRange_B} $A_1^+$: Scattering length and effective range computed from the linear interpolation between levels 1 and 2, and parameters for the position of the $D_{s0}^*(2317)$ bound state $m_B$ derived from the requirement $\cot\delta (p_{B})=\I$. The second uncertainty given for values in MeV corresponds to the uncertainty in the lattice scale $a$. The experimental value of $m_K+m_{D}-m_{B}$ is averaged over $D^+K^0$ and $D^0K^+$ thresholds.  }
\end{table*}

\begin{table*}[tbp]
\begin{center}
\begin{ruledtabular}
\begin{tabular}{c|cccccl|lccccc}
level  & $t_0$ & basis &$\textrm{fit}\atop\textrm{range}$       &$\textrm{fit}\atop\textrm{type}$  & $\tfrac{\chi^2}{d.o.f}$ & $Ea$  & $E-\bar m~$& $(a p)^2$  & $a p ~\cot(\delta)$ &$p^2$       & $ p ~\cot(\delta)$ \vspace{-3pt} \\
         &          &          & &    &                                    &            &           [MeV]      &           &          & [GeV$^2$]      & [GeV]         \\              
\hline                                
\multicolumn{6}{l}{Ensemble (1)}\\
\hline
1 & 3 & $O_{1,4,7-11}$& 10-15 & 1exp$^c$ & 0.12 & 1.3340(28) & 377.4(4.2)& -0.0382(11) & -0.1701(44) & -0.0970(29) & -0.2709(69) \\
2 & 3 & $O_{1,4,7-11}$& 10-15 & 1exp$^c$ & 1.45 & 1.3761(75) & 444(12)&  &  &  &  \\
3 & 3 & $O_{1,4,7-11}$& 10-15 & 1exp$^c$ & 0.50 & 1.4645(38) & 585.3(5.9) &  0.0314(17) & -0.1998(101)& 0.0796(44) & -0.318(16) \\
4 & 3 & $O_{1,4,7-11}$&  4-11 & 2exp$^c$ & 0.54 & 1.6681(80) & 909(13) &  0.1707(52) & -1.09(38)& 0.4330(132)& -1.73(60) \\
\hline
\multicolumn{6}{l}{Ensemble (2)}\\
\hline
1 & 3 & $O_{1,4,7-11}$      & 4-14 & 2exp$^c$ & 1.58 & 1.0260(52) & 392(11)  & -0.0097(19) & -0.086(14) & -0.0460(88) & -0.188(30) \\
2 & 3 & $O_{1,4,7-11}$      & 4-11 & 2exp$^c$ & 1.00 & 1.0791(47) & 507(10)  &     &    &    &    \\
3 & 3 & $O_{1,4,7-11}$      & 4-11 & 2exp$^c$ & 0.71 & 1.0811(64) & 511(14)  &  0.0106(26) & -0.071(25) &  0.050(12) & -0.155(54) \\
4 & 3 & $O_{1,4,7-11}$      & 4-11 & 2exp$^c$ & 0.45 & 1.1723(93) & 710(20)  &  0.0506(45) & -0.113(116)&  0.239(21) & -0.24(25) \\
\hline
1 & 3 & $O_{1,2,4,5,9,11}$   & 4-20 & 2exp$^c$ & 0.27 & 1.0259(35) & 391.3(7.6)  & -0.0098(13) & -0.0867(99)  & -0.0463(63)  & -0.189(22) \\
2 & 3 & $O_{1,2,4,5,9,11}$   & 4-12 & 2exp$^c$ & 0.89 & 1.0765(34) & 501.3(7.4)  &     &    &    &    \\
3 & 3 & $O_{1,2,4,5,9,11}$   & 4-12 & 2exp$^c$ & 1.80 & 1.0799(24) & 508.7(5.2)  &  0.0101(11) & -0.0762(103) &  0.0478(50)  & -0.166(22) \\
4 & 3 & $O_{1,2,4,5,9,11}$   & 4-12 & 2exp$^c$ & 1.27 & 1.162(18)  & 688(40)     &  0.0458(85) & -0.28(55)&  0.217(40) & -0.6(1.2) \\
\end{tabular}
\end{ruledtabular}
\end{center}
\caption{ Energy levels for irrep $T_1^+$ for both ensembles and $s$-wave phase shifts extracted from them (time-slices start from t=0 such that $t_0=3$ corresponds to the fourth time-slice). The superscript $c$ indicates a correlated fit and $\bar m=\tfrac{1}{4}(m_{D_s}+3 m_{D_s^*})$. For ensemble (2) we show the fit result for two sets of interpolators to point out the possible systematic error due to the choice.  The second level is identified with $D_{s1}(2536)$ coupling weakly to $s$-wave (see the discussion in the text); we therefore do not include it in the phase shift analysis. }\label{tab:T1}
\end{table*}

Some results for this channel have already been presented in Ref. \cite{Mohler:2013rwa} and therefore we will be brief here. 
We analyzed the contribution of  the various interpolators to the energy eigenstates by (a) the overlap factors $\langle n|O_i\rangle$ and the eigenvectors and (b) by determining the eigenstates considering subsets of the complete set. Fig. \ref{fig:A1_levels_subsets} shows the impact of the $DK$ operators  on the determination of lowest  eigenstates. In this plot all time fit ranges for the 2-exponential fits are 3-10 and the results are compatible with 1-exponential fits in the range 7-10. We only show the lowest energy levels where a clear plateau behavior of the effective energies is observed. Our final results have used different fit ranges chosen optimally for the basis used. Table \ref{tab:A1} gives the energy values for the eigenstates using the complete operator basis for both ensembles. 

\begin{figure}[tbp]
\begin{center}
\includegraphics*[width=0.42\textwidth,clip]{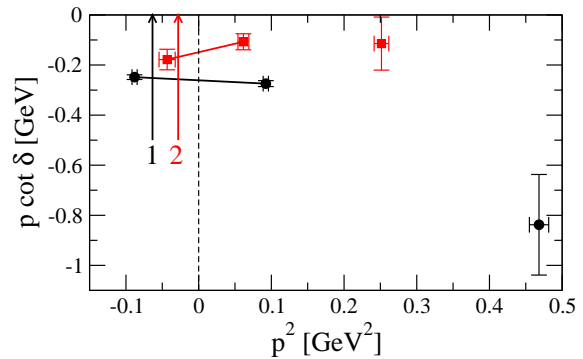}
\end{center}
\caption{Effective range fits for $A_1^+$, cf. Table \ref{tab:A1_EffRange_B}.
Ensemble (1) black dots, ensemble (2) red squares; 
the vertical arrows give the positions of the bound state for ensembles (1) and (2), see Table \ref{tab:A1_EffRange_B}, the dashed line indicates the threshold.}
\label{fig:A1_eff_range_fits}
\end{figure}

As discussed in more detail in Ref. \cite{Mohler:2013rwa} we identify the lowest eigenstate as lying close to the bound state $D_{s0}^*(2317)$ and the level above threshold with the lowest scattering state. With L\"uscher's relation (see Sect. \ref{sec:eff_range}) we determine values of $\mathrm{Re}(T^{-1})$ and therefrom values of the scattering length and the effective range (Table \ref{tab:A1_EffRange_B} and Fig. \ref{fig:A1_eff_range_fits}). Our results are compatible with the analysis in Ref. \cite{Liu:2012zya} where the authors performed a lattice calculation in a variety of other channels and extracted the relevant low-energy constants of the chiral effective field theory. These low-energy constants were then used to predict the $DK$ ($I = 0$) scattering length indirectly.

From the bound state condition \eq{eq:B-condition} we obtain the location of the $D_{s0}^*(2317)$ bound state given in Table \ref{tab:A1_EffRange_B}. The resulting mass is shown together with other channels in Fig. \ref{fig:summary}. 

\subsection{$D_{s1}$}

\begin{figure}[tb]
\begin{center}
\includegraphics*[width=0.45\textwidth,clip]{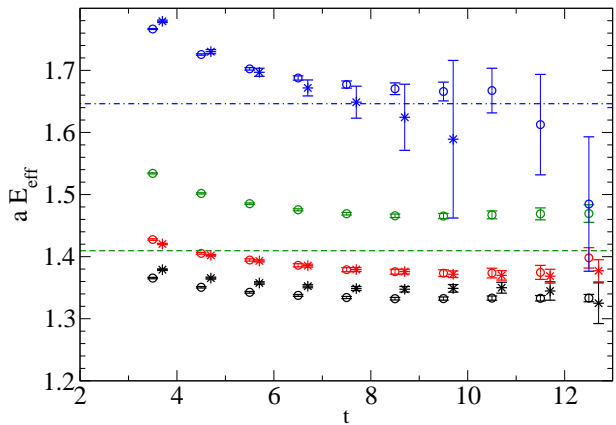}
\end{center}
\caption{Ensemble (1), $T_1^+ $ with $t_0=3$: Effective energies of the lowest four energy levels. We compare the results obtained including the $D^*K$ operators (circles, $O_{1-11}$) with the results obtained without  those (stars, $O_{1-8}$).The horizontal broken lines in the upper plot indicate the positions of $D^*(0)K(0)$ and $D^*(1)K(-1 )$ in the non-interacting case. Note the ``missing  state'' in the second case. }
\label{fig:ensemble1_eff_energies}
\end{figure}

\begin{figure}[tbhp]
\begin{center}
\includegraphics*[width=0.47\textwidth,clip]{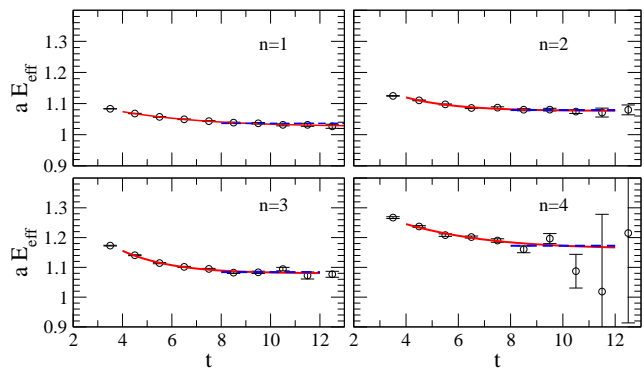}
\end{center}
\caption{Ensemble (2),  $T_1^+ (O_{1,2,4,5,9,11})$: Effective energies of the lowest four energy levels; the fits are to the eigenvalues but here we show the result in the effective energy plots:   2-exponential fits (red) and 1-exponential fits (broken blue line) are  consistent. }
\label{fig:ensemble2_eff_energies_a}
\end{figure}

\begin{figure*}[tbhp]
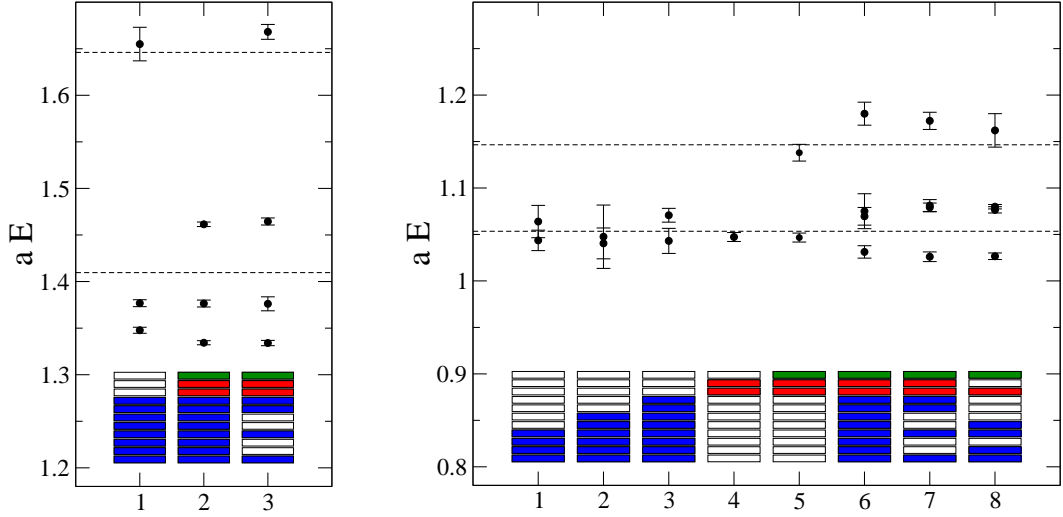

\begin{center}
\includegraphics*[height=6.8cm,clip]{levels_tz2_ens_1.eps}\hspace*{24pt}
\includegraphics*[height=6.8cm,clip]{levels_tz2_4-11_a.eps}
\end{center}
\caption{$T_1^+ $: Effective energies as obtained for various subsets of operators for ensemble (1) (lhs) and (2) (rhs). The fit type and fit range is shown in Table \ref{tab:T1}. The horizontal broken lines indicate the positions of $D^*(0)K(0)$ and $D^*(1)K(-1 )$ in the non-interacting case. The boxes indicate the operators considered in each case (blue: $\qbar q$ , red:  $D^*(0)K(0)$, green: $D^*(1)K(-1 )$. }
\label{fig:T1_diff_sets}
\end{figure*}

\begin{table}[htbp]
\begin{center}
\begin{ruledtabular}
\begin{tabular}{l|cc}                               
set &$m_{D_{s1}(2536)}\!-\!\tfrac{1}{4}(m_{D_s}\!+\!3 m_{D_s^*})$& $m_{D_{s1}(2536)}\!-\!m_K\! -\! m_{D^*}$ \\
&  [MeV]  & [MeV] \\ 
 \hline 
\multicolumn{3}{l}{Ensemble (1)}\\
\hline     
& 444(12) & -53(12) \\              
\hline                                 
\multicolumn{3}{l}{Ensemble (2)}\\
\hline                                   
set 1& 507(10)& 56(11) \\                         
set 2& 501(8) & 50(8) \\                  
\hline                               
\multicolumn{3}{l}{Experiment}\\
\hline                                
 &459  &31\\                             
\end{tabular}
\end{ruledtabular}
\end{center}
\caption{\label{tab:T1_Ds2536}Comparison of the mass of  $D_{s1}(2536)$ with experiment.
}
\end{table}
For the two ensembles we had to rely on slightly different fit ranges. For ensemble (1) we used 1-exponential fits, for ensemble (2) we  used 2-exponential fits (and checked consistency with 1-exponential fits). The final results are summarized in Table \ref{tab:T1}.  For ensemble (2) we show the fit result for two sets of interpolators to point out the possible systematic error due to that choice. 

Figures \ref{fig:ensemble1_eff_energies} and  \ref{fig:ensemble2_eff_energies_a} demonstrate the typical behavior of the 
effective energies for the two ensembles. For ensemble (1) the highest level (actually the third for that set of operators) has a plateau-like signal only when including the $D^*K$ interpolators. In Fig. \ref{fig:ensemble2_eff_energies_a} we also plot the results of the 2- and 1-exponential fits (the errors of the asymptotic values are given in Table \ref{tab:T1}).

In Fig. \ref{fig:T1_diff_sets} we give an overview on the energy levels resulting from different subsets of interpolators
in the variational analysis. One clearly sees that including the $D^*(0)K(0)$ interpolators 9 and 10 introduces new levels. 
In ensemble (1) the signal for the 4th level is too noisy, when considering all 11 interpolators, but is clearly seen
for the subset $1,4,7-11$. 

From ensemble (2) the effect is even more apparent: Allowing for only the $D^*K$ interpolators one finds energies very close
to the non-interacting case. When coupling all interpolators one finds level shifts due to interaction.
For this ensemble the 2nd and 3rd level are very close when considering all types of interpolators, whereas in ensemble (1) these
are well separated. This supports the observation that only one of the levels is dominated by $D^*K$.

\subsubsection{Interpretation of the energy levels}

The lowest level is identified with the experimental state $D_{s1}(2460)$, below $D^*K$ threshold. It couples to $D^*K$ in $s$-wave even in the heavy quark ($m_c\to\infty$) limit \cite{Isgur:1991wq}. The level is seen already for $\qbar q$  interpolators alone but it is down-shifted by about 20 MeV (ensemble (1)) or 33 MeV (ensemble (2)) if the $D^*(0)K(0)$ interpolators are included.

The second state in both ensembles is identified with $D_{s1}(2536)$. In ensemble (1) with the heavier Pion the state lies below $m_D^*+m_K$, but in the ensemble (2) we find it above this threshold. The mass of  $D_{s1}(2536)$ is given ``naively'' from the 2nd energy level in Table \ref{tab:T1}  and compared with experiment in Table \ref{tab:T1_Ds2536}.
 
In the heavy quark limit, according to Ref. \cite{Isgur:1991wq} $D_{s1}(2536)$ does not couple to $D^*K$ in $s$-wave. We find that the composition of the states with regard to the $\qbar q$ operators is fairly independent of whether the $D^*K$ operators are included or not. This can be seen by the eigenvector components as well as the overlap factors $\langle n=2|O_{1-8}\rangle$. The level is not seen if only $D^*K$ interpolator are used. Experimentally the state is above $D^*K$ threshold but has - in spite of this - the very small decay width $\Gamma\simeq 0.92~$MeV; coupling in $s$- and in $d$-wave is observed. The experiment gives  $g\simeq 0.2~$GeV (for a total width $\Gamma\equiv g^2 p/s$) which indeed seems $m_c$ suppressed in comparison to $g[D_1(2430)\to D^*\pi]\simeq 2~$GeV.  
So it is reasonable to assume that the coupling $D_{s1}(2536)\to D^* K$ in $s$-wave is indeed small. Due to the small coupling the ``avoided level crossing'' region is so narrow that we may treat this state as decoupled from the $D^*K$ scattering channel. L\"uscher's equation for $\delta_0$ then does not affect this energy level. For this reason the corresponding value of $p ~\cot\delta_0$ is not provided in Table \ref{tab:T1}. 

Level three is dominated by $D^*(0) K(0)$ as can be seen by prevailing $\langle n=3|O_{9-10}\rangle$ and analogously the 4th level is dominated by $D^*(1) K(-1)$.

\begin{table*}[htbp]
\begin{center}
\begin{ruledtabular}
\begin{tabular}{l|cc|cccc}                               
set & $a_0^{D^*K}$ & $r_0^{D^*K}$ & $(a p_{B})^2$ & $a m_{B}$ & $m_K+m_{D^*}-m_{B}$  &$m_{B}-\tfrac{1}{4}(m_{D_s}+3 m_{D_s^*}) $ \\
& [fm] & [fm]& &    & [MeV] & [MeV] \\ 
 \hline 
\multicolumn{3}{l}{Ensemble (1)}\\
\hline     
& -0.665(25) & -0.106(37) & -0.0301(15) & 1.3511(35) & 93.2(4.7)(1.0) & 404.6(4.5)(4.2) \\              
\hline                                 
\multicolumn{3}{l}{Ensemble (2)}\\
\hline                                   
set 1& -1.15(19) & 0.13(22) & -0.0071(22) & 1.0336(60) & 43.2(13.8)(0.6)& 408(13)(5.8) \\                         
set 2& -1.11(11) & 0.10(10) & -0.0073(16) & 1.0331(41) & 44.2(9.9)(0.6) & 407.0(8.8)(5.8) \\                  
\hline                               
\multicolumn{3}{l}{Experiment}\\
\hline                                
 &&       &             &                      & 44.7   &  383\\                             
\end{tabular}
\end{ruledtabular}
\end{center}
\caption{\label{tab:T1_EffRange_B}
 $T_1^+$  Scattering length and effective range computed from the linear interpolation between levels 1 and 3, and parameters for the position of the $D_{s1}(2460)$ bound state $m_B$ derived from the requirement $\cot\delta (p_{B})=\I$. The second uncertainty given for values in MeV corresponds to the uncertainty in the lattice scale $a$. The experimental value of $m_K+m_{D^*}-m_{B}$ is averaged over $D^{*+}K^0$ and $D^{*0}K^+$ thresholds. 
}
\end{table*}

\begin{figure}[htbp]
\begin{center}
\includegraphics*[width=0.42\textwidth,clip]{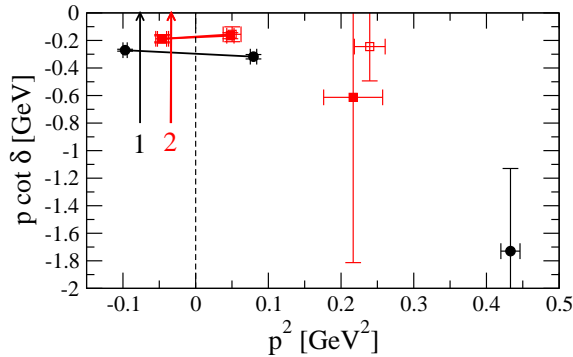}
\end{center}
\caption{Effective range fits for $T_1^+$, cf. Table \ref{tab:T1_EffRange_B}.
Ensemble (1): black dots, ensemble (2): red open/full (set 1/2) squares. The vertical arrows show the positions of the bound state, see Table \ref{tab:T1_EffRange_B}, the dashed line indicates the threshold.}
\label{fig:T1_eff_range_fits}
\end{figure}

\begin{table}[htbp]
\begin{center}
\begin{ruledtabular}
\begin{tabular}{c|cccccc|cc}
$n$  & $t_0$ & basis &$\textrm{fit}\atop\textrm{range}$&$\textrm{fit}\atop\textrm{type}$ & $\tfrac{\chi^2}{d.o.f}$ & $Ea$  & $E-\bar m~$ \\
         &          &          & &&  &                                         &  [MeV]                 \\              
\hline    
\multicolumn{3}{l}{Ensemble (1)}\\
\hline                                
1 &  2 & $O_{1,2}$      &  11-17 & 1exp$^c$ & 0.28 & 1.3939(64) & 473(10)(5)\\
\hline                            
\multicolumn{3}{l}{Ensemble (2)}\\
 \hline
1 &  2 & $O_{1,2}$      &  3-14  & 2exp$^c$ & 1.06 & 1.0852(35) & 520(8)(7) \\
 \hline
\multicolumn{3}{l}{Experiment}\\
 \hline
 \multicolumn{3}{l}{$D_{s2}^*(2573)$}    &&&& & 496 \\
 \end{tabular}
\end{ruledtabular}
\end{center}
\caption{ Ground state energy for $T_2^+$. The superscript $c$ indicates a correlated fit and $\bar m=\tfrac{1}{4}(m_{D_s}+3 m_{D_s^*})$ is the spin-averaged $D_s$ meson mass. The second uncertainty given for values in MeV corresponds to the uncertainty in the lattice scale $a$.}\label{tab:T2}
\end{table}

\begin{figure*}[t]
\begin{center}
\includegraphics*[width=0.7\textwidth,clip]{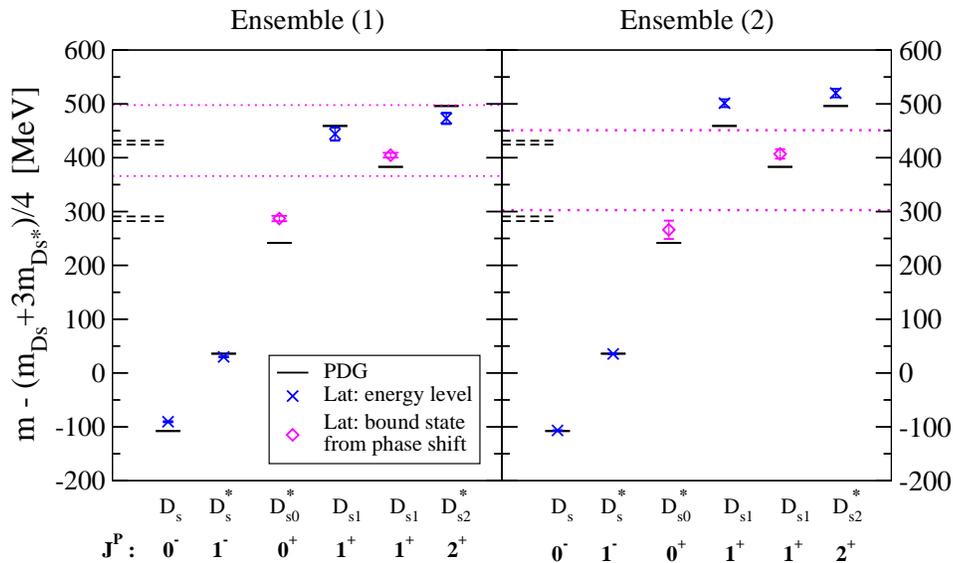}
\end{center}
\caption{ Resulting $D_s$ spectrum for all channels. The masses are presented with respect to spin-averaged mass $\tfrac{1}{4}(m_{D_s}+3m_{D_s^*})$. The diamonds and crosses display our lattice results, while black full lines correspond to experiment.  The magenta diamonds show masses extracted via phase shift analysis and correspond to the pole position in the $T-$matrix. Masses extracted as energy levels in a finite box are displayed as blue crosses. The dotted lines correspond to $DK$ and $D^*K$ lattice thresholds, while dashed lines on the left of each pane are thresholds in experiment.   }
\label{fig:summary}
\end{figure*}

\subsubsection{Bound state and threshold behavior}

As discussed in Sect. \ref{sec:eff_range}  we can use the values of $p\cot\delta(p)$  from L\"uscher's relation \eq{eq:luescher_z} to determine the effective range parametrization near threshold. From levels 1 and 3 we find the values in Table \ref{tab:T1_EffRange_B}. The pole condition $\delta(p_{B})=\I$ renders the pole for the $D_{s1}(2460)$ bound state (B) with parameters given in Table \ref{tab:T1_EffRange_B} as well. The resulting mass is shown together with other channels in Fig. \ref{fig:summary}.

\subsection{$D_{s2}^*$}

Experiments observe the decay of $D_{s2}^*(2573)\to DK$ (with a width of 17(4) MeV). Since the mass of this state is quite far away from the first (in $d$-wave) relevant level $D(1)K(-1)$ we did not include the $DK$ interpolators. This also applies to a possible $D^*K^*$  contribution. We find the energy levels shown in Table \ref{tab:T2}.  Identifying this level ``naively'' with  the $D_{s2}^*(2573)$ gives mass differences also shown in the table compared to the experimental value.

\section{Summary and conclusions}\label{sec:conclusions}

Lattice QCD is used to simulate $DK$ and $D^*K$ scattering in order to study the positive parity charmed strange mesons $D_{s0}^{*}(2317)$, $D_{s1}(2460)$, $D_{s1}(2536)$ and $D_{s2}^*(2573)$. These mesons are interesting from a physics point of view for two main reasons. First, the masses of the scalar and axial vector mesons are close to their charm light partners even though the strange quark is much heavier than the light. Second, contrary to the expectation from quark models, the $D_{s0}^{*}(2317)$ and $D_{s1}(2460)$ are both narrow below-threshold states. Many models and lattice QCD studies attempted to understand this. In lattice calculations, a combination of unphysical thresholds and treatment within the single hadron approach rendered the masses too high. In particular, the effects of $DK$ and $D^*K$ thresholds were not taken into account explicitly. 

In our simulation we include $DK$ and $D^*K$ scattering operators. We work with two quite different ensembles of gauge configurations: ensemble (1) with $N_f=2$ dynamical fermions and $m_\pi\simeq 266$ MeV and ensemble (2) with $N_f=2+1$ dynamical fermions and $m_\pi\simeq 156$ MeV. The necessary correlators involve backtracking quark loops and the calculation is made feasible by using the standard distillation and - on large lattices - the stochastic distillation method.

 We determine the low lying energy spectrum from which the scattering amplitude near threshold is derived via L\"uscher's finite volume method. This method allows us to study successfully the threshold parameters and near threshold resonance and bound states. We extract the binding momenta and the masses of the below-threshold bound states $D_{s0}^{*}(2317)$ and $D_{s1}(2460)$. The final mass spectrum is compiled in Fig.~\ref{fig:summary} for both ensembles.

$J^P=0^+$-channel: The $D_{s0}^{*}(2317)$ with $J^P=0^+$ benefited most from the inclusion of scattering operators; the level assigned to it in the single hadron approach was just slightly above threshold and when $DK$ scattering operators were included it decoupled into two states, one attributed to the scattering channel and the other to the physical bound state. The analytical continuation of the scattering amplitude combined with L\"uscher's finite volume method  allowed us to establish the existence of a below threshold state with binding energy $37(17)~$MeV which is compatible with the $D_{s0}^{*}(2317)$ and which we therefore identify with the $D_{s0}^{*}(2317)$.

$J^P=1^+$-channel: The $D_{s1}(2460)$ with $J^P=1^+$ appeared below threshold even in the single hadron approach. However, the inclusion of $D^*K$ scattering operators significantly improved the signal and the detailed analysis showed that $D_{s1}(2460)$ indeed has a considerable four-quark component. Repeating a similar analysis as for the scalar channel, we find the binding energy $44(10)~$MeV of $D_{s1}(2460)$ in agreement with experiment. We also find the narrow $D_{s1}(2536)$, which is above threshold for ensemble (2) with Pion masses close to physical. Experiments find this state in $d-$wave and $s-$wave, while the $s$-wave coupling is expected to disappear in the $m_c\to \infty$ limit. 

$J^P=2^+$-channel: Here we did not include $DK$ interpolators as the energy of the first such interpolator is far above the lowest energy state. The mass of $2^{+}$ state $D_{s2}^*(2573)$,  obtained using just $\bar qq$ interpolators, is also presented in Fig.~\ref{fig:summary}.

Comparing the two ensembles, the overall agreement with the observed $D_s$ spectrum in Fig.~\ref{fig:summary}  improves for the ensemble with almost physical Pion mass. Unlike in previous studies, an unambiguous signal for the $D_{s0}^{*}(2317)$ and $D_{s1}(2460)$ as strong interaction bound states below the $DK$ and $D^*K$ thresholds is obtained. To achieve this, close to physical quark masses and the inclusion of $DK$ and $D^*K$ operators in the basis of lattice interpolating fields were crucial ingredients. 

\begin{appendix}
\section{Interpolators}\label{app_a}

In this study quark-antiquark interpolating fields of the type $O_i^{\sbar c}=\sbar A_i c$  as well as meson-meson interpolators are used. All interpolators are projected to total momentum zero. The operators are irreducible representations of the  octahedral group $O_h$.
 
The  quark-antiquark interpolator kernels are given in Table \ref{tab:interpolators} for the three cases
with $J^P=0^+$ (irrep $A_1^+$), with $J^P=T^+$ (irrep $T_1^+$), and
with $J^P=2^+$ (irrep $T_2^+$),

For the cases $J^P=0^+$ (irrep $A_1^+)$ and $J^P=1^+$ (irrep $T_1^+$) we also included  meson-meson interpolators in $s$-wave.  The mesons are projected to $\vec p$ individually, the total momentum is zero.

For $J^P=0^+$ (irrep $A_1^+)$ we use $DK$:
\begin{align}
\label{eq:Op_DK_A1}
O_1^{DK}&=\left[\bar{s}\gamma_5u\right](\vec p=0)\left[\bar{u}\gamma_5c\right](\vec p=0)+\left\{u\rightarrow d\right\}\;,\nonumber\\
O_2^{DK}&=\left[\bar{s}\gamma_t\gamma_5u\right](\vec p=0)\left[\bar{u}\gamma_t\gamma_5c\right](\vec p=0)+\left\{u\rightarrow d\right\}\;,\nonumber\\
O_3^{DK}&=\!\!\!\!\!\!\!\!\!\sum_{\vec p=\pm e_{x,y,z}~2\pi/L}\!\!\!\!\!\!\!\left[\bar{s}\gamma_5u\right](\vec p)\left[\bar{u}\gamma_5c\right](-\vec p)+\left\{u\rightarrow d\right\}\;.
\end{align}

For  $J^P=1^+$ (irrep $T_1^+$) we use $D^*K$:
 \begin{align}
\label{eq:Op_DK_T1}
O_{1,k}^{D^*K}&=\left[\bar{s}\gamma_5 u\right](\vec p=0)\left[\bar{u}\gamma_k c\right](\vec p=0)+\left\{u\rightarrow d\right\}\;,\nonumber\\
O_{2,k}^{D^*K}&=\left[\bar{s}\gamma_t\gamma_5 u\right](\vec p=0)\left[\bar{u}\gamma_t\gamma_k c\right](\vec p=0)+\left\{u\rightarrow d\right\}\;,\nonumber\\
O_{3,k}^{D^*K}&=\!\!\!\!\!\!\!\!\!\sum_{\vec p=\pm e_{x,y,z}~2\pi/L}\!\!\!\!\!\!\!\left[\bar{s}\gamma_5 u\right](\vec p)\left[\bar{u}\gamma_k c\right](-\vec p)+\left\{u\rightarrow d\right\}\;.
\end{align}
The index $k$ denotes the polarization.

\begin{table}[t]
\begin{ruledtabular}
\begin{tabular}{c|c|c|c}
\T\B  Lattice & Quantum numbers  & Interpolator & Operator \\
\T\B  irrep &  $J^{PC}$ in irrep & label & \\
 \hline
$A_1^{+}$ & $0^{+}$, $4^{+}$, $\dots$ & 1 & $\bar{q}q^\prime$ \\
 & & 2 & $\bar{q}\gamma_i{\overrightarrow{\nabla}_i}q^\prime$ \\
 & & 3 & $\bar{q}\gamma_t\gamma_i{\overrightarrow{\nabla}_i}q^\prime$ \\
 & & 4 & $\bar{q}{\overleftarrow{\nabla}_i}{\overrightarrow{\nabla}_i}q^\prime$ \\
\hline
 $T_1^{+}$ & $1^{+}$, $3^{+}$, $4^{+}$, $\dots$ & 1 & $\bar{q}\gamma_i\gamma_5q^\prime$ \\
 & & 2 & $\bar{q}\epsilon_{ijk}\gamma_j{\overrightarrow{\nabla}_k}q^\prime$\\
 & & 3 & $\bar{q}\epsilon_{ijk}\gamma_t\gamma_j{\overrightarrow{\nabla}_k}q^\prime$\\
 & & 4 & $\bar{q}\gamma_t\gamma_i\gamma_5q^\prime$\\
 & & 5 & $\bar{q}\gamma_5{\overrightarrow{\nabla}_i}q^\prime$\\
 & & 6 & $\bar{q}\gamma_t\gamma_5{\overrightarrow{\nabla}_i}q^\prime$\\
 & & 7 & $\bar{q}{\overleftarrow{\nabla}_i}\gamma_j\gamma_5{\overrightarrow{\nabla}_i}q^\prime$ \\
 & & 8 & $\bar{q}{\overleftarrow{\nabla}_i}\gamma_t\gamma_j\gamma_5{\overrightarrow{\nabla}_i}q^\prime$\\
\hline
 $T_2^{+}$ & $2^{+}$, $3^{+}$, $4^{+}$, $\dots$ & 1 & $\bar{q}|\epsilon_{ijk}|\gamma_j{\overrightarrow{\nabla}_k}q^\prime$\\
 & & 2 & $\bar{q}|\epsilon_{ijk}|\gamma_t\gamma_j{\overrightarrow{\nabla}_k}q^\prime$\\
\end{tabular}
\end{ruledtabular}
\caption{\label{tab:interpolators}Table of $\sbar c$ interpolators used for $D_s$ mesons; in addition we use $DK$ and $D^*K$  interpolators for irreps $A_1^+$ and $T_1^+$. Interpolators are sorted by irreducible representation of the octahedral group $O_h$ and by the parity  quantum number $P$. The operators $\nabla_k$ indicate covariant lattice derivatives. The reduced lattice symmetry implies an infinite number of continuum spins in each irreducible representation of $O_h$. The Dirac matrix for the time direction is denoted by $\gamma_t$.}
\end{table}

\end{appendix}
\acknowledgments
We thank Anna Hasenfratz and the PACS-CS collaboration for providing gauge configurations and Martin L\"uscher for making his DD-HMC software available. D.~M. would like to thank E.~Eichten, F.-K. Guo, M.~Hansen, A.~Kronfeld, Y.~Liu and J.~Simone for insightful discussions. The calculations were performed on computing clusters at TRIUMF, the University of Graz and at Jozef Stefan Institute. This work is supported in part by the Austrian Science Fund (FWF):[I1313-N27], by the Slovenian Research Agency ARRS project N1-0020 and by the Natural Sciences and Engineering Research Council of Canada. Fermilab is operated by Fermi Research Alliance, LLC under Contract No. De-AC02-07CH11359 with the United States Department of Energy. Special thanks to the Institute for Nuclear Theory (University of Washington) for hospitality.

\clearpage

\bibliographystyle{h-physrev4}
\bibliography{Lgt,newbib}

\end{document}